%% file: main.tex
\documentclass[5p,times]{elsarticle}

\usepackage{graphicx}
\usepackage{amssymb}
\usepackage{booktabs}
\usepackage{array}
\usepackage{microtype}
\usepackage[htt]{hyphenat}
\usepackage{xcolor}
\usepackage{comment}
\usepackage{url}
\usepackage[colorlinks=true,linkcolor=blue,citecolor=blue,urlcolor=blue]{hyperref}

\journal{}
\makeatletter
\def\ps@pprintTitle{\let\@oddhead\@empty\let\@evenhead\@empty\let\@oddfoot\@empty\let\@evenfoot\@empty}
\makeatother

\begin{document}

\begin{frontmatter}

\title{PegasusAI Studio: Legible Cyberinfrastructure and Accountable AI Agents
for Scientific Workflows}

\author[unc]{Komal Thareja}
\author[usc]{Mats Rynge}
\author[umass]{Berent Aldikacti}
\author[utk]{Angelina Ju}
\author[utk]{Ian Lumsden}
\author[unc]{Anirban Mandal}
\author[usc]{Rajiv Mayani}
\author[utk]{Kin Wai Ng}
\author[usc]{Hamza Safri}
\author[unc]{Michael Stealey}
\author[utk]{Sai Swaminathan}
\author[utk]{Michela Taufer}
\author[usc]{Karan Vahi}
\author[umass]{Michael Zink}
\author[usc]{Ewa Deelman}
\affiliation[unc]{organization={University of North Carolina at Chapel Hill}, city={Chapel Hill}, state={NC}, country={USA}}
\affiliation[usc]{organization={University of Southern California}, city={Los Angeles}, state={CA}, country={USA}}
\affiliation[utk]{organization={University of Tennessee, Knoxville}, city={Knoxville}, state={TN}, country={USA}}
\affiliation[umass]{organization={University of Massachusetts Amherst}, city={Amherst}, state={MA}, country={USA}}

\begin{abstract}
AI coding agents can now construct scientific workflows, but operating them on
shared cyberinfrastructure (CI) raises problems that code generation does not
solve. Agents need current knowledge of the workflow system and its execution
sites and access to the runtime state of distributed executions, while
facilities need to attribute every agent action to a user and bound it by that
user's privileges. We present PegasusAI Studio, an AI-assisted science gateway
built on the Pegasus and HTCondor stack. Studio makes CI legible to agents by
projecting one facility-curated knowledge base into the native configuration
of several coding agents, by introspecting workflow interfaces and execution
sites, and by grounding diagnosis in workflow and scheduler telemetry. Agent
actions operate within a trust envelope that combines federated
authentication, per-user execution environments and scheduler credentials,
provider-neutral model access, observability, and approval controls for state-changing
actions. The same application runs as a JupyterHub-based gateway on a research
cloud and as an Open OnDemand application on HPC systems. We evaluate Studio
through case studies of legacy CASA weather workflow modernization, a
gravitational-wave reanalysis, an EEG analysis built with an open-weight
model, and an Event Horizon Telescope (EHT) M87 image reconstruction that recovered from two failed runs to a
zero-failure execution. 
The evaluation establishes that these mechanisms operate as designed.

\end{abstract}


\end{frontmatter}

\input{sections/01-introduction-studio}
\input{sections/02-background-studio}
\input{sections/03-related-work-studio}
\input{sections/07-architecture}
\input{sections/08-knowledge-projection}
\input{sections/09-lifecycle}
\input{sections/10-trust}
\input{sections/11-evaluation-studio}
\input{sections/12-discussion}
\input{sections/13-threats-studio}
\input{sections/14-artifact-studio}
\input{sections/15-conclusion-studio}

\section*{Declaration of competing interest}
The authors declare that they have no known competing financial interests or
personal relationships that could have appeared to influence the work
reported in this paper.

\section*{Acknowledgments}
This work is supported by Pegasus AI NSF grant \#2513101 and the FABRIC Testbed (NSF \#2330891).

\section*{Data availability}
The Studio application, Open OnDemand package, canonical knowledge base, and
case-study workflow artifacts are available at the locations described in
Section~\ref{sec:artifact}.

\section*{Declaration of generative AI and AI-assisted technologies in the
writing process}
During the preparation of this work the authors used Claude Code
(Anthropic) to draft manuscript text from an author-written specification
and to typeset tables and figures. After using this tool, the authors
reviewed and edited the content as needed and take full responsibility for
the content of the published article.

\bibliographystyle{elsarticle-num}
\bibliography{refs}

\end{document}

%% file: sections/01-introduction-studio.tex
\section{Introduction}
\label{sec:introduction}

Most large scientific computations now run on shared cyberinfrastructure (CI):
campus clusters, national allocations, research clouds, and high-throughput
pools that many users and projects share. Workflow management systems make
that infrastructure usable. Pegasus~\cite{deelman2015pegasus} and
HTCondor~\cite{thain2005condor}, for example, run production workloads across
many scientific domains, handling data staging, retries, recovery, and
provenance so that a scientist does not have to. AI coding agents can now
write substantial scientific software, including the workflows these systems
execute. Writing the workflow, however, is only part of the work.

Shared CI is a heterogeneous execution environment in which computation,
data, software, and services are distributed across resources and sites.
Operating an agent there requires three things a researcher's laptop does not
supply. First, the agent needs current knowledge of workflow-system and site
conventions. Second, it needs access to the runtime state of distributed executions, which lives in the
workflow database and the scheduler rather than in a terminal. And finally, the facility needs an authorization model that attributes every agent action to
a user and bounds it by that user's privileges.

Our companion study of AI-assisted workflow
construction~\cite{pegasus_method_companion} found that workflow logic was
usually not the dominant source of effort. Across a twelve-workflow corpus,
much of the effort instead went into runtime and configuration debugging
involving containers, schedulers, data movement, software environments, and
remote execution. 
This echoes earlier experience integrating an Earth science workflow end to end across distributed cyberinfrastructures, where similar integration challenges dominated~\cite{DBLP:conf/ucc/RoaROVMGKGHRLVL23}.
That finding motivated the system described here.

We present PegasusAI Studio, an AI-assisted science gateway built on the
Pegasus and HTCondor stack. Its design follows two principles. First, CI must
be made \emph{legible} to AI agents, by exposing system knowledge,
operational conventions, interfaces, execution sites, and runtime state in
machine-consumable form. Second, AI agents must be made \emph{accountable}
to the CI, by placing their actions inside the same identity, authorization,
and auditing mechanisms that govern human users.

Studio realizes legibility through three mechanisms. Knowledge projection
compiles one facility-curated knowledge base into the native configuration
of several coding agents. Structural and site introspection exposes workflow
interfaces and execution sites. Telemetry grounding gives the agent workflow
and scheduler state in place of terminal output. Studio realizes
accountability through a \emph{trust envelope} that combines federated
authentication, per-user execution environments, per-user scheduler tokens,
provider-neutral model access, observability, and user approval controls for
state-changing actions. 
The same Studio application image can be deployed in two environments: as a JupyterHub-based science gateway on a research cloud or as an Open OnDemand application on an HPC cluster.

We evaluate Studio through case studies drawn from real workflow development
on the deployed gateway: the modernization of three legacy CASA radar
workflows, a gravitational-wave reanalysis, an EEG group analysis built with
an open-weight model, and a reconstruction of the Event Horizon Telescope (EHT) M87 imaging pipelines~\cite{ehtIV2019, patel2023eht} that recovered from two failed runs to a
zero-failure execution. The evidence is observational. It shows each
mechanism operating as designed, and it does not yet quantify how much
effort the mechanisms save.

The main contributions are:
\begin{enumerate}
\item \textbf{An end-to-end architecture for AI-assisted workflow operation
on shared CI}, built on the Pegasus and HTCondor stack without modifying either
and organized around mechanisms that are applicable to other workflow design and execution 
systems (Section~\ref{sec:architecture}).
\item \textbf{Knowledge projection}, which compiles one curated body of
workflow and site knowledge into the native configuration of multiple
coding agents (Section~\ref{sec:knowledge}).
\item \textbf{A telemetry-grounded workflow lifecycle}, in which agents
diagnose runs from structured workflow and scheduler state rather than from
terminal output (Section~\ref{sec:lifecycle}).
\item \textbf{A trust envelope for accountable agents}, which ties each
agent action to the user's identity and authorization and provides user-approval controls over state-changing operations (Section~\ref{sec:trust}).
\item \textbf{Deployment evidence} from workflows built and run inside
Studio, including CASA nowcasting modernization, gravitational-wave
reanalysis, EEG analysis, and the EHT M87 reconstruction
(Section~\ref{sec:evaluation}), with the application and knowledge base
released as open source (Section~\ref{sec:artifact}).
\end{enumerate}

%% file: sections/02-background-studio.tex
\section{Background and System Context}
\label{sec:background}

\subsection{Scientific workflows and the execution stack}
Pegasus~\cite{deelman2015pegasus} represents a scientific computation as a
resource-independent directed acyclic graph (DAG) and plans it onto concrete execution sites, inserting
data staging, container handling, retries, and provenance capture. Studio uses
Pegasus without modification and submits workloads through HTCondor~\cite{thain2005condor}, a system designed to manage high-throughput job execution across distributed resources. 
HTCondor's per-user identity tokens, hold reasons, and execution state are important because they provide both accountability and diagnostic information.

Pegasus and HTCondor provide a production-proven execution foundation for
large-scale scientific computing. Pegasus has been used to execute workflows
with up to millions of tasks across heterogeneous distributed resources,
including clusters, national CI, and clouds, with mechanisms
for data management, failure recovery, provenance, and portable execution
across environments~\cite{deelman2015pegasus}. HTCondor has similarly evolved
over decades of production use in distributed high-throughput computing and
has been demonstrated at scales of hundreds of thousands of simultaneously
running jobs across geographically distributed
resources~\cite{thain2005condor,fajardo2015htcondor}. PegasusAI Studio therefore does not
reimplement scheduling, data movement, retries, recovery, or distributed
execution in the agent layer. Instead, it builds on these established
capabilities and focuses the AI layer on workflow construction, operation, and
diagnosis.

\subsection{Science gateways}
A science gateway gives a community browser-based access to computation, data,
and domain tools~\cite{lawrence2015gateways}. Studio builds on two common
patterns: JupyterHub~\cite{kluyver2016jupyter}, which authenticates users and
provisions per-user environments, and Open OnDemand~\cite{hudak2018openondemand},
which provides browser access to HPC resources. CILogon~\cite{basney2014cilogon}
provides federated identity for the gateway deployment.

\subsection{AI coding agents and workflow construction context}
Modern coding agents iteratively plan, invoke tools, observe results, and revise
actions under a permission model. The Model Context Protocol
(MCP)~\cite{mcp2024} exposes documentation and service tools to agents. Studio
integrates the PegasusAI construction toolchain evaluated separately in the
companion methodology paper~\cite{pegasus_method_companion}: documentation
grounding, reusable skills, spec-first interaction, an iterative development
lifecycle, and codified workflow failure-mode knowledge.

This paper takes that construction methodology as given and asks a different
question: what systems mechanisms are needed to operate AI-assisted workflow
development on shared CI? Studio provides the knowledge
distribution, runtime visibility, identity, authorization, and gateway
environment for that setting.

%% file: sections/03-related-work-studio.tex
\section{Related Work}
\label{sec:related}
Workflow systems such as Pegasus~\cite{deelman2015pegasus},
Nextflow~\cite{ditommaso2017nextflow}, Snakemake~\cite{molder2021snakemake},
CWL~\cite{crusoe2022cwl}, Parsl~\cite{babuji2019parsl}, and
Galaxy~\cite{galaxy2022} provide portable and reproducible execution, but generally assume the user can
already express a workflow in the system's interfaces. Science gateways such
as Open OnDemand~\cite{hudak2018openondemand}, Galaxy,
HUBzero~\cite{mclennan2010hubzero}, and JupyterHub~\cite{kluyver2016jupyter}
provide mature web access to
shared computation and solve multi-user concerns such as authentication and
resource isolation, but do not by themselves give AI agents workflow-system
knowledge, runtime telemetry, or an accountable execution identity.

Recent work has begun to connect AI assistants and agents to research-computing
environments. Science-gateway efforts have explored AI-assisted user support
and the broader role of AI in gateway-based research computing~\cite{biggs2023aigateways,biggs2024hpcassistant,sperhac2024aigateways}, while other
work has connected LLM agents to HPC resources through parallel execution
frameworks such as Parsl~\cite{ma2025llmhpc}. These systems demonstrate that AI assistants can
operate alongside or invoke shared computing resources. PegasusAI Studio
focuses on a complementary systems problem: integrating the agent with the
workflow lifecycle while exposing structured execution state and preserving
the identity, authorization, and audit mechanisms of the underlying facility.

AI coding assistants and scientific
agents~\cite{boiko2023coscientist,bran2024chemcrow} provide code generation,
chat, and tool use, while retrieval-augmented generation can ground models in current
documentation. Existing integrations are typically editor- or application-
scoped. Studio differs by integrating the agent with the workflow lifecycle,
execution telemetry, gateway identity, and resource-provider controls rather
than only with source code. The companion methodology paper addresses the
separate question of how agents construct workflows from scientific intent and
existing artifacts~\cite{pegasus_method_companion}.

%% file: sections/07-architecture.tex
\section{PegasusAI Studio: Architecture}
\label{sec:architecture}

The PegasusAI workflow-construction methodology was initially developed in a command-line
setting, where one operator configured the agent context, worked from a local
shell, and diagnosed remote execution failures through logs and terminal
output. Although adequate for studying workflow construction, this
setting did not give the agent direct access to the mechanisms 
shared facilities use to distribute site knowledge, expose execution state,
and attribute actions to users. This limitation became particularly important
because the companion evaluation found that
much of the remaining
workflow-construction effort arose from runtime and
configuration problems. Studio addresses this gap by moving the
same construction methodology into a facility-managed environment that exposes
CI knowledge and runtime state while preserving facility identity and
authorization controls.

Studio must support two directions of integration. CI must expose system
knowledge and execution state in forms agents can use, while agent actions
must remain inside the identity, authorization, and auditing mechanisms of
the facility. We call these requirements \emph{legibility} and
\emph{accountability}, respectively. Table~\ref{tab:twoway} maps them to the corresponding
Studio mechanisms, and the following sections describe them.

\begin{table}[t]
\centering
\caption{Studio mechanisms that connect AI agents to cyberinfrastructure, expose cyberinfrastructure state to agents, and govern agent actions.}
\label{tab:twoway}
\small
\begin{tabular}{@{}>{\raggedright\arraybackslash}p{0.24\linewidth}>{\raggedright\arraybackslash}p{0.70\linewidth}@{}}
\toprule
Direction & Mechanism in Studio \\
\midrule
Agent access to CI & Chat on a coding-agent engine with specialist agents and the
Pegasus tools reachable over MCP, agent-enabled terminals, and skills embedded
in the workflow IDE. A JupyterLab sidebar chat aware of the notebook the
user has open. \\ \hline
CI state to agent & The knowledge projection layer
(Section~\ref{sec:knowledge}). Telemetry grounding
(Section~\ref{sec:lifecycle}). Argparse introspection. Per-workflow
placement on the shared pool or the user's own annex, and an administrator's
site catalog under Open OnDemand (OOD)~\cite{hudak2018openondemand}. MCP documentation servers. \\ \hline
CI governance of agent & Federated identity, per-user tokens, approval posture, a
provider-neutral model gateway with a prompt guard, and per-user usage
telemetry with consent-gated content capture (Section~\ref{sec:trust}). \\
\bottomrule
\end{tabular}
\end{table}

\subsection{Design constraints that shaped the system}
\label{sec:constraints}

Three constraints recur throughout the design, and every architectural
choice below can be traced to one of them. First, the workflow stack is
unmodified: Studio integrates Pegasus and HTCondor only through their standard
interfaces. This is deliberate. Rather than reimplementing scheduling, data
movement, recovery, provenance, and distributed execution in the AI layer,
Studio relies on the mature capabilities of the underlying workflow and
scheduling systems. This keeps the approach portable to existing
installations and keeps the trust argument of Section~\ref{sec:trust} grounded in
mechanisms providers already accept. Second, AI assistance must remain optional and degrade gracefully when unavailable: 
a user with no model key still gets a full
workflow environment, and an institution can disable or substitute any
agent. Third, the system must operate behind arbitrary reverse-proxy
prefixes, since gateway sessions live under per-user URL paths. This last
constraint sounds minor but consumed a disproportionate share of
engineering effort. We return to it in Section~\ref{sec:discussion}.

\subsection{Two tiers}
\label{sec:tiers}

\begin{figure*}[t]
\centering
\includegraphics[width=0.90\textwidth]{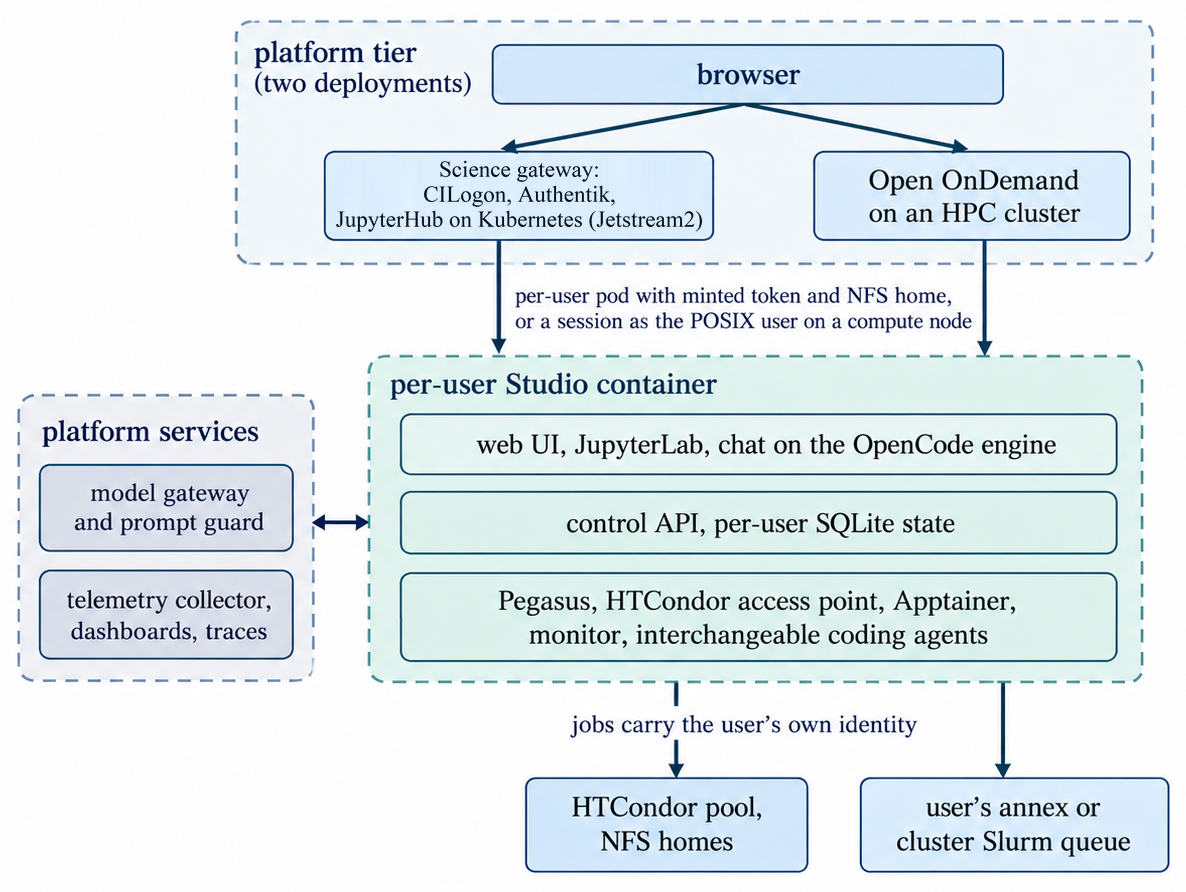}
\caption{The two tiers and the two platform deployments. One container image
holds a complete single-user workflow environment. The platform tier that
spawns and authenticates it is either the JupyterHub gateway on a
research cloud or an OOD app on a campus cluster. Every process
in the container, agents included, runs as the user's own identity with
the user's scheduler token, over a Pegasus and HTCondor stack.
Model access goes through a provider-neutral gateway with a prompt guard,
and usage telemetry flows to a collector inside the platform.}
\label{fig:architecture}
\end{figure*}

Studio has two tiers: a self-contained application container that gives one
user a complete workflow development environment, and a platform tier
that spawns, authenticates, and provisions one container instance per user for a community
of users. The split is deliberate. The container runs anywhere a single user
can run Docker~\cite{merkel2014docker} or Apptainer~\cite{kurtzer2017singularity}, supporting laptop and lab-host use, while
the platform adds identity and shared compute without changing the
image. Figure~\ref{fig:architecture} shows both tiers.

The platform tier supports two deployment models. The first is the public science
gateway described in Section~\ref{sec:platform}: JupyterHub on a
Kubernetes~\cite{burns2016kubernetes} cluster, with federated login, one pod per user, and a shared
HTCondor pool. The second is an OOD
application for sites that already run OOD on an HPC cluster. For this deployment, the
same application is packaged in an Apptainer image on a Rocky
Linux base, so that the cluster's own Slurm~\cite{yoo2003slurm} client tools are bound into the container, and a
standard Batch Connect app launches it as the logged-in POSIX user on a
compute node. Each session gets its own Studio home directory
rather than the user's real home, internal services use
per-session port because Apptainer shares the host network namespace, and
a session token gates the front proxy. Workflow jobs reach the cluster's
Slurm queue through an HTCondor access point. The OOD deployment was
validated end to end with three test users on a three-node Slurm cluster on the FABRIC testbed~\cite{baldin2019fabric}. A head node runs the Slurm
controller, the HTCondor access point, and the OOD portal, and two compute
nodes run Slurm.
A workflow built in a session ran to completion with all 12 of its jobs
executed as Slurm jobs. This deployment has not yet been exercised at a production site.
In both deployment models, the platform changes how the container is
launched, authenticated, and connected to compute resources, but not the
application environment inside the container.

\subsubsection{The Studio container}
\label{sec:container}

\begin{figure*}[t]
\centering
\includegraphics[width=0.9\textwidth]{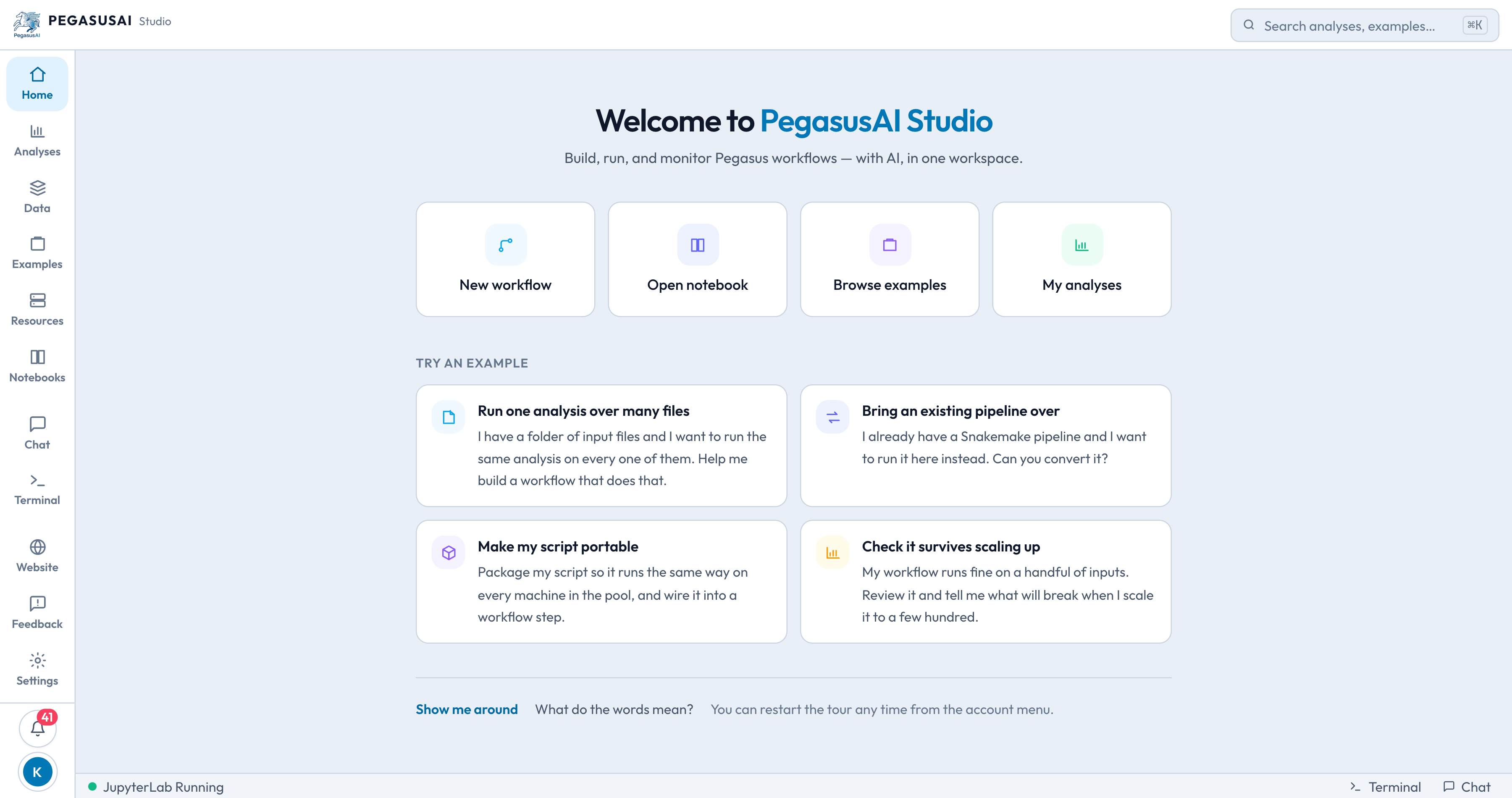}
\caption{The Studio home page. The navigation rail on the left lists the
views in the order a scientist works (home, analyses, data, examples,
resources, notebooks), followed by the chat, terminal, website, feedback, and
settings entries. The home page offers four starting actions and four
example requests that each hand a prompt to the assistant, and a link that
starts the guided tour.}
\label{fig:studio-home}
\end{figure*}

A single \texttt{linux/amd64} image bundles four kinds of components behind
one reverse proxy:

\begin{itemize}
\item \textbf{User interface.} A web application whose views follow the
  order in which a scientist works: a home page that hands a prompt to
  the assistant, an analyses page for workflow projects and runs with a
  live graph, a data page for uploads, an examples page that lists
  community workflows from tagged public repositories, notebooks,
  a resources page for choosing where workflows run, and settings. A
  guided tour introduces the interface, and a single Run button prepares,
  builds, plans, and submits a workflow. The chat surface is also embedded
  in JupyterLab as a sidebar extension that knows which notebook the
  user currently has open and can read and edit its cells. Figures~\ref{fig:studio-home}, \ref{fig:studio-form}, and~\ref{fig:studio-dag}
  show the current design.
\item \textbf{Control API.} A service that coordinates application state and actions:
  workflow scanning, planning and submission, run monitoring, container
  image builds, AI tool installation and configuration, chat engine
  dispatch, and notifications. State lives in a per-user SQLite database
  inside the user's workspace, such that application state has no external
  service dependencies.
\item \textbf{Workflow stack.} The container runs Pegasus WMS, an
  HTCondor access point, and Apptainer for containerized jobs. Because
  these are the standard components, the user plans and submits workflows
  through the same interfaces used at a conventional Pegasus site. A
  fourth component, \texttt{workflow-monitor}, is a small read-only
  service that follows the Pegasus workflow database and scheduler state
  and converts them into the structured event stream that both the user
  interface and the agents consume (Section~\ref{sec:lifecycle}). 
\item \textbf{AI agents.} Several interchangeable coding agents, at the time
  of writing Claude Code~\cite{tool-claudecode} (Anthropic),
  OpenCode~\cite{tool-opencode} and Crush~\cite{tool-crush} (both open
  source), Codex CLI~\cite{tool-codex} (OpenAI), and
  Antigravity~\cite{tool-antigravity} (Google), installed
  system-wide. The web chat is a client of the OpenCode engine running in
  the user's own container, with the Pegasus tools (workflow scanning,
  planning, submission, run diagnosis) exposed to it over MCP. In its
  default mode, the chat routes each message to one of the specialist
  agent roles of Section~\ref{sec:knowledge}, so a failure report reaches
  the workflow debugger and a design request reaches the workflow
  architect. The agents are ordinary third-party tools. Studio configures them (Section~\ref{sec:knowledge}).
\end{itemize}

The container keeps no user accounts of its own. Whatever spawns it
(JupyterHub, OOD, or a plain \texttt{docker run}) is responsible
for identifying the user, and the container runs as the UID it is given. 
Access to the application is then gated at the front proxy using the
authentication mechanism provided by the deployment environment. Under JupyterHub,
every request (pages, API, terminal, and JupyterLab) is checked against the
hub through its OAuth flow, and only the pod's owner or a JupyterHub administrator is admitted. Under OOD a per-session token gates the
proxy. By keeping user identity and multi-user authentication outside the
application image, the same container can run in a local single-user setting
or inside a federated multi-user gateway. The platform tier delegates
authentication, user isolation, and session management to mature components
such as JupyterHub, CILogon, and Open OnDemand rather than reimplementing those
capabilities inside Studio.

\subsubsection{The gateway platform}
\label{sec:platform}

\begin{figure*}[t]
\centering
\includegraphics[width=0.9\textwidth]{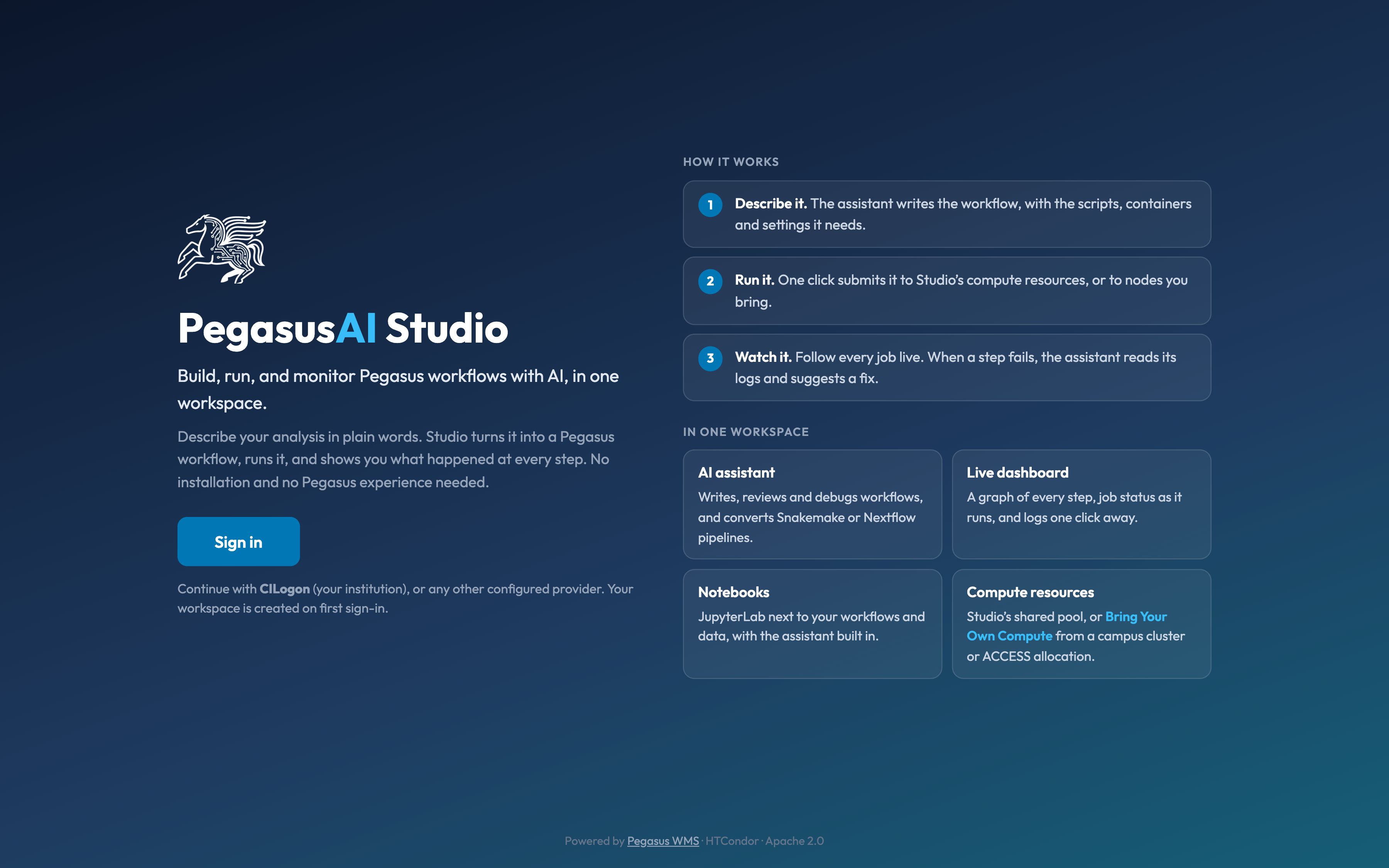}
\caption{The public gateway's sign-in page. A user signs in through CILogon
with their home institution's credentials, and the workspace is created on
first sign-in. The page summarizes the three-step lifecycle (describe, run,
watch) and the parts of the workspace.}
\label{fig:studio-login}
\end{figure*}

The gateway platform is the multi-user tier: the components that
authenticate users and provision one Studio container per user. The public
deployment runs on a self-managed Kubernetes cluster on the Jetstream2
research cloud~\cite{hancock2021jetstream2}. Users authenticate through
CILogon~\cite{basney2014cilogon} with their home institution credentials,
with Authentik brokering that identity to JupyterHub (Figure~\ref{fig:studio-login}). A hook in the
JupyterHub spawner, the component that creates a user's pod, runs before
each session. It creates the user's home directory on a shared NFS server
if absent, mints a fresh HTCondor identity token scoped to that user into
that home directory, and mounts the home into the pod. Because the home
outlives the pod and the token is re-minted at every spawn, the user's pod
starts with persistent storage and a valid scheduler credential, and every job
the user (or an agent acting on behalf of the user) submits carries the
user's own identity into the HTCondor pool (Section~\ref{sec:trust}).

Compute follows the same per-user model. The HTCondor pool has a central
access point and a set of execution points. Execution points include
cloud instances, which a small controller service scales with demand, and 
privileged pods inside the cluster. The pool also flocks to the OSPool~\cite{pordes2007osg} for users who
supply an OSPool project. A user who needs more than the shared pool can lease nodes,
including GPU nodes, from any Slurm clusters on which they hold an allocation,
through HTCondor's annex mechanism (Section~\ref{sec:trust}). Three further platform services sit
beside the pool: a model gateway and prompt guard, a telemetry collector
with its dashboards, and an in-cluster trace store, all described in
Section~\ref{sec:trust}. GitOps tooling reconciles the entire platform
from a git repository, so the deployment itself is versioned and
reviewable.

%% file: sections/08-knowledge-projection.tex
\section{Knowledge Projection: Making CI Legible to Agents}
\label{sec:knowledge}

Before Studio was introduced, in the command-line phase, workflow and failure-mode knowledge was
maintained in context files configured for individual projects, users, and
agent tools~\cite{pegasus_method_companion}. That arrangement made the
quality of an agent's context depend on who had configured it and allowed
copies to diverge as tools and users multiplied. Studio instead treats this
knowledge as facility-managed infrastructure: it is curated, maintained centrally, versioned
with the platform, and projected into each supported agent's native
configuration for every user. 

\subsection{One canonical knowledge base}

The knowledge base is maintained in a single canonical open-source repository, the PegasusAI
plugin marketplace~\cite{pegasusai-plugin}, 
which contains the
skills, agent roles, templates, and documentation-server definitions described
below. 
The Studio container
checks it out at a pinned revision. Studio keeps no second copy: its
continuous-integration build fails if one reappears in the application
repository, because an earlier arrangement with two hand-maintained
copies led them to diverge (one copy taught Docker while the
other taught Apptainer). The checkout retains its version history, so an
administrator can advance or roll back the knowledge inside a running
container without rebuilding the application image. The repository contains four kinds of
artifacts:

\begin{itemize}
\item \textbf{Skills.} Procedural expertise for recurring tasks, each a
  canonical markdown document with structured metadata: scaffold a workflow
  from a workflow description, convert a Snakemake or Nextflow workflow,
  generate a wrapper script for a workflow step, produce an Apptainer
  definition for a tool stack, review a workflow for known pitfalls,
  diagnose a failed run, answer questions from the documentation and code
  examples, and provision execution resources. These are the same skills as
  the companion skill catalog~\cite{pegasus_method_companion}, now curated by the facility rather than the
  user.
\item \textbf{Agent roles.} Supported coding agents can delegate longer tasks to
specialized agent instances with their own instructions and context. 
An agent role is the instruction set for one such
  specialized agent, designed for a
longer-running task,  for example a workflow architect
  that designs a DAG, a data engineer that writes fetch and harmonize
  steps, a workflow debugger that works through a failed run. The web
  chat's default mode routes each message to one of these roles.
\item \textbf{Templates and examples.} Complete, working workflow
  generators (for example, a fan-out and fan-in over many inputs, or a
  fetch-train-predict workflow) and configuration templates (site catalogs
  for the Studio pool and for ACCESS~\cite{accesspegasus}, container definitions, resource
  profiles). Skills point the agent to these files so that implementation guidance can be grounded in executable examples rather than prose alone.
\item \textbf{Documentation servers.} A registry of Model Context Protocol
  (MCP)~\cite{mcp2024} endpoints exposing the documentation of Pegasus and
  of Kiso~\cite{kiso} as searchable tools, so agents retrieve current reference
  material instead of relying on training data. Kiso is an experiment-management tool for provisioning HTCondor pools on cloud
and edge testbeds and is exposed through the \texttt{/kiso} skills ~\cite{pegasus_method_companion}.
\end{itemize}

The result is a single versioned body of operational knowledge maintained
by the workflow-system and gateway administrators and projected consistently
across users and agent tools.

\subsection{Per-tool adapters}

Each supported agent tool reads its configuration in a different format
and from different files. An \emph{adapter} is a small program that
translates the canonical knowledge base into one tool's format. When a
user enables an AI tool, its adapter compiles the canonical knowledge
into that tool's native configuration in the user's workspace: a
\texttt{CLAUDE.md} and \texttt{.mcp.json} for Claude Code, an
\texttt{AGENTS.md} with \texttt{opencode.json} and skill and agent
directories for OpenCode, an \texttt{AGENTS.md} for Codex, a context file
for Antigravity, and the equivalent for Crush, as well as configurations for Studio's own JupyterLab sidebar and the web chat.
Adapters write to the user's home directory so the configuration is global
to the user's session rather than tied to one project checkout, and
re-running an adapter reconciles the output with the current knowledge
version. The same adapter also carries per-user settings that must reach
the agent, such as the telemetry consent of Section~\ref{sec:trust}. 
When such a setting changes, Studio restarts the agent engine so that the
updated configuration takes effect immediately rather than waiting for a new
session.

The pattern is one-to-many by construction: $N$ supported tools require $N$
thin adapters over one knowledge base, and supporting a new agent means
writing one additional adapter to that tool's configuration interface rather
than maintaining a separate copy of the knowledge base. In our experience each adapter is on the
order of a hundred lines of code, because 
the substantive work of deciding and maintaining what an agent needs to know about the workflow system is centralized in the canonical layer.

\subsection{Structural legibility: interface introspection}

\begin{figure*}[t]
\centering
\includegraphics[width=0.8\textwidth]{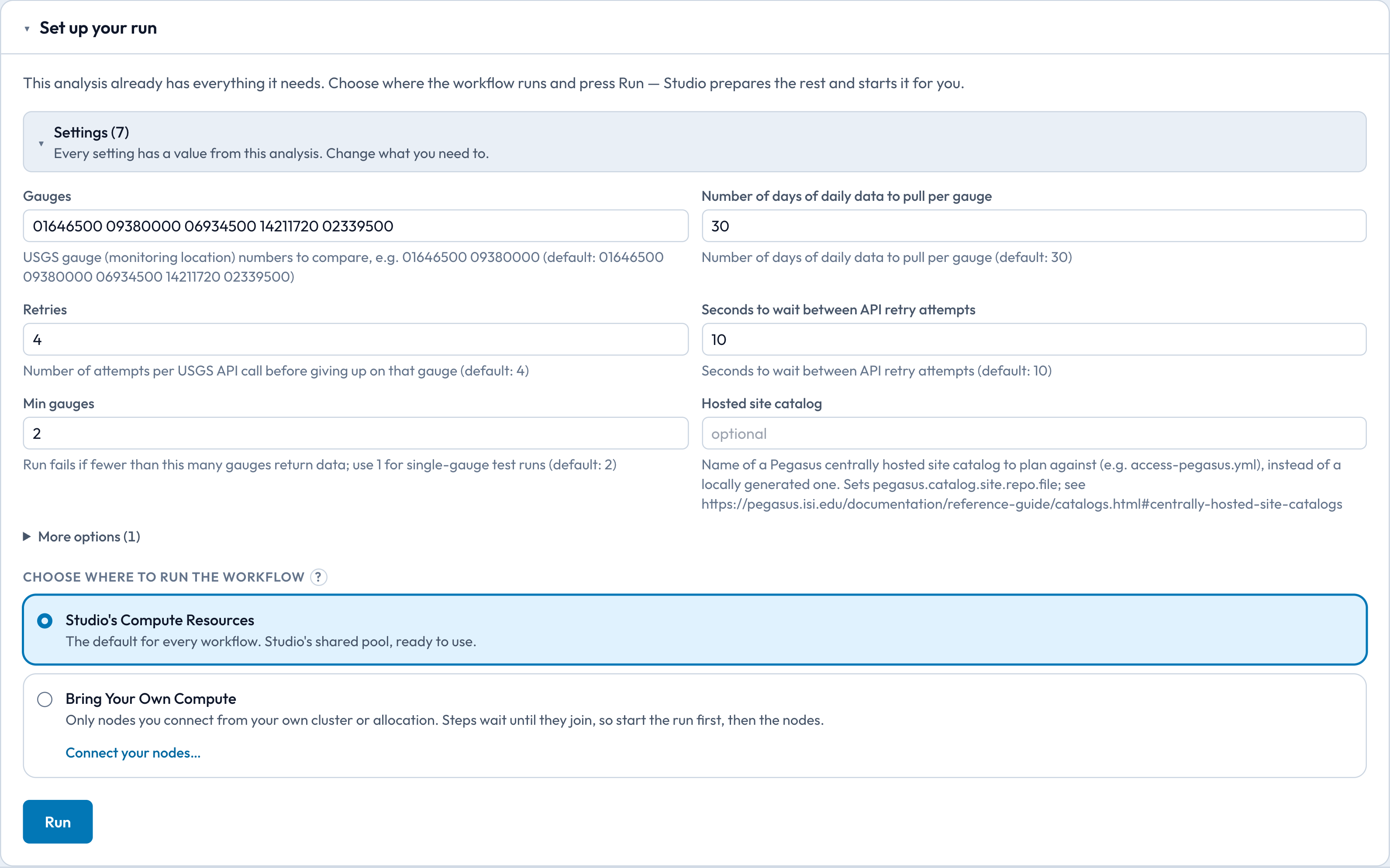}
\caption{The run panel of a workflow project in Studio, here for a
streamflow workflow the assistant built from one chat prompt. The settings
(gauge list, lookback window, retry policy, minimum number of gauges, and an
optional site catalog) are derived from the generator's argparse interface by
abstract-syntax-tree introspection, without executing user code. Every field,
default, and help string comes from the interface the agent wrote. Below the
settings the user chooses where the workflow runs (Studio's shared pool, or
only nodes the user connects from their own cluster or allocation). One
button prepares, plans, and submits the workflow.}
\label{fig:studio-form}
\end{figure*}

Legibility extends beyond documents into program structure. Every Pegasus
workflow project in Studio exposes a generator script whose command-line
interface defines the workflow's parameters. Studio parses that interface by
analyzing the script's abstract syntax tree, without executing user code,
and derives both the parameter form shown to the user and the parameter
schema given to the agent. The same mechanism means an agent-authored
generator can become immediately operable from the UI: if the agent writes a
conventional argparse interface, the platform can run it. A convention
thereby becomes a contract that binds both sides: the platform knows how to operate generators
that follow the convention, and the knowledge base
instructs every agent to follow it. Figure~\ref{fig:studio-form} shows the
result: a workflow project page whose parameter form is derived from
the generator's own interface. The contract has been exercised on
agent-authored work: the gravitational-wave case of
Section~\ref{sec:evaluation} was launched from the derived form, the
first launch surfaced an interface defect at exactly that boundary, and
the fixed workflow ran to completion from the form.

\subsection{Site legibility: one execution site}

A third kind of legibility concerns where a workflow runs. The Pegasus
convention used by Studio generators is that a generator
describes only the \texttt{local} site (submit-host scratch and output
storage) and plans against a single execution site named
\texttt{compute}, whose definition comes from a site catalog. Studio includes
that definition, so a generated workflow never names a cluster. On the
gateway, \texttt{compute} is Studio's own pool. Extra capacity comes from
the user's HTCondor annex on a cluster they hold an allocation on
(Section~\ref{sec:trust}) or from flocking to the OSPool, not from a
different site. The run form then offers two placements per workflow:
everywhere except the annex, which is the default, or only the user's annex
(Figure~\ref{fig:studio-form}). Studio records the choice as a scheduler
requirement in the workflow's own Pegasus properties, so a
\texttt{pegasus-plan} typed in a terminal places jobs exactly as the Run
button does. Under Open OnDemand, an administrator supplies the site catalog
for the host cluster, and the same \texttt{compute} name points at it. The
agent plans against exactly the site the deployment defines, so the local
policy of Section~\ref{sec:why-projection} (preferred sites, allocations)
is expressed once rather than in every generated workflow.

\subsection{Why projection rather than per-tool authoring}
\label{sec:why-projection}

The projection design rests on three expected benefits: 
(1)  It should keep expertise
current, because a fix to a skill (for example, a newly learned rule from
the companion failure-mode table~\cite{pegasus_method_companion} about how fetch jobs must handle missing credentials
on execute nodes) reaches every tool on the next reconciliation. (2) It should keep
expertise consistent, because two agents given the same task draw on
the same curated guidance, which matters when users switch tools mid-project. And (3), 
it should lower the barrier for institutions, because an execution site can encode local
policy (preferred sites, container registries, data placement rules) in one
place and project that guidance into every supported agent. The OpenCode projection has been
exercised end to end: an open-weight model working from the compiled
knowledge base scaffolded, debugged, and executed a new EEG analysis
workflow to completion on the gateway, providing an end-to-end test of the projected knowledge
with a different agent and model stack in Section~\ref{sec:evaluation}.

%% file: sections/09-lifecycle.tex
\section{The Telemetry-Grounded Lifecycle}
\label{sec:lifecycle}

\begin{figure}[t]
\centering
\includegraphics[width=\columnwidth]{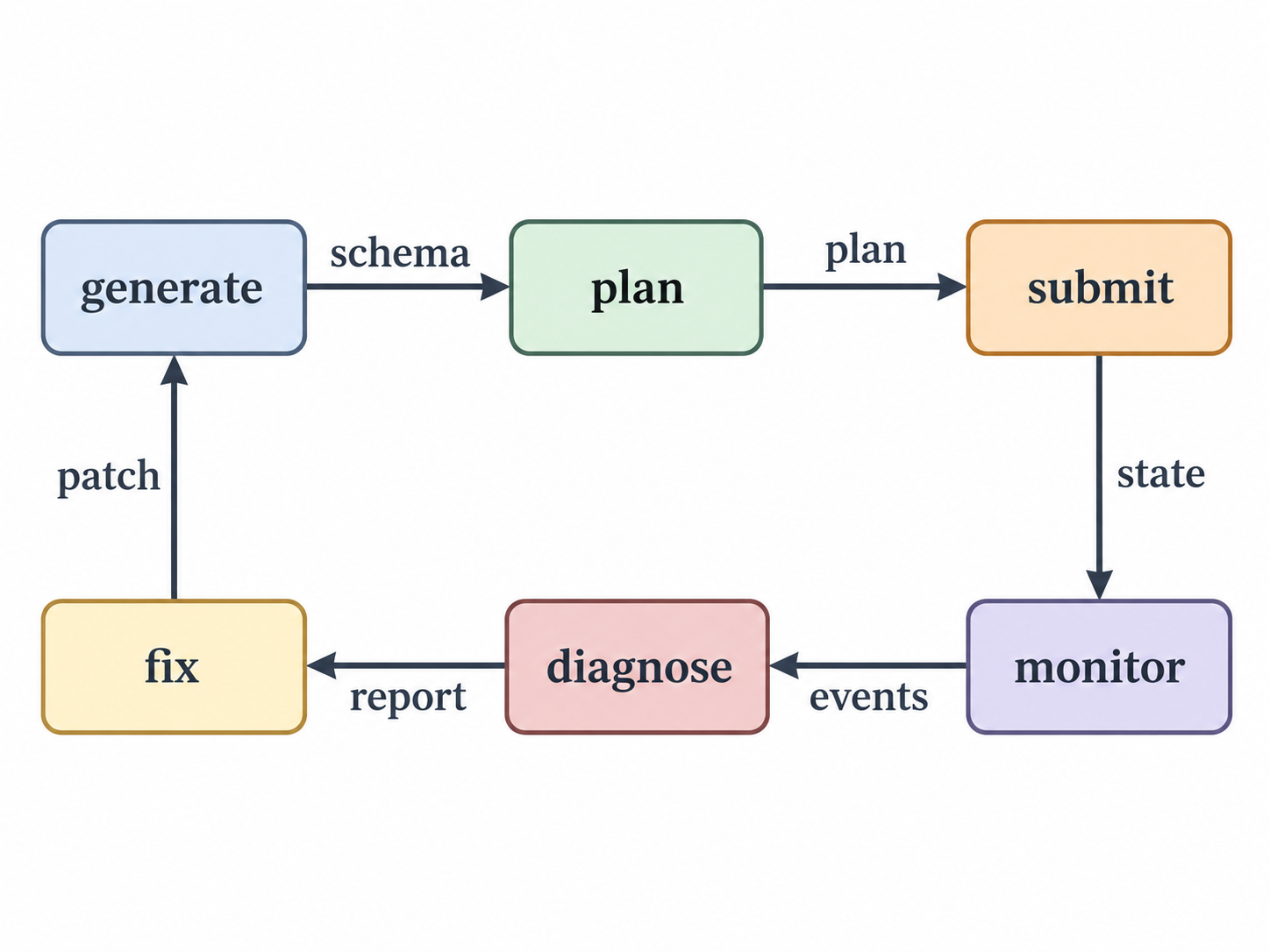}
\caption{The workflow lifecycle as the unit of integration. Each transition 
makes structured information from the workflow and execution systems available
to Studio, including planner output, scheduler state, the JSONL event stream,
and analyzer reports, together with the introspected \texttt{argparse} schema
derived from the workflow generator. Proposed fixes return to the workflow as
patches, closing the diagnose, fix, and resubmit lifecycle through replanning and
resubmission rather than relying on terminal scraping.
}
\label{fig:lifecycle}
\end{figure}

The companion methodology evaluation found that much of the remaining
workflow-construction effort arose during runtime and configuration debugging~\cite{pegasus_method_companion}. In the command-line setting, the agent saw
only the execution information surfaced through terminal output and manually
opened logs, while the authoritative state remained in the workflow database
and scheduler. Studio therefore treats the workflow lifecycle, rather than the
editor, as the unit of integration. Figure~\ref{fig:lifecycle} shows the lifecycle and the structured
information exchanged between its stages.

\subsection{Stage by stage}

\textbf{Generate.} A workflow starts from a community example (fetched
live from tagged public repositories), a skill-driven scaffold, or existing
user code. The generator's introspected interface
(Section~\ref{sec:knowledge}) yields the parameter form and schema.

\textbf{Plan and submit.} Studio invokes the standard Pegasus planner and
records the run. Before planning, it checks the workflow's configuration
against the target site. A workflow written for a cluster with a shared
filesystem can fail when run on an HTCondor pool whose workers do not share a filesystem. In that case, jobs may fail without an error
that clearly identifies the configuration mismatch.  Studio therefore stops such a run
before submission, with a message naming both the setting and the
deployment. This is a recurring failure mode encoded directly in the platform rather than left only
as agent knowledge. Nothing about planning is AI-specific, which is the point:
the agent operates the same tools as the users, so its actions are
reviewable in the same terms.

\begin{figure*}[t]
\centering
\includegraphics[width=0.9\textwidth]{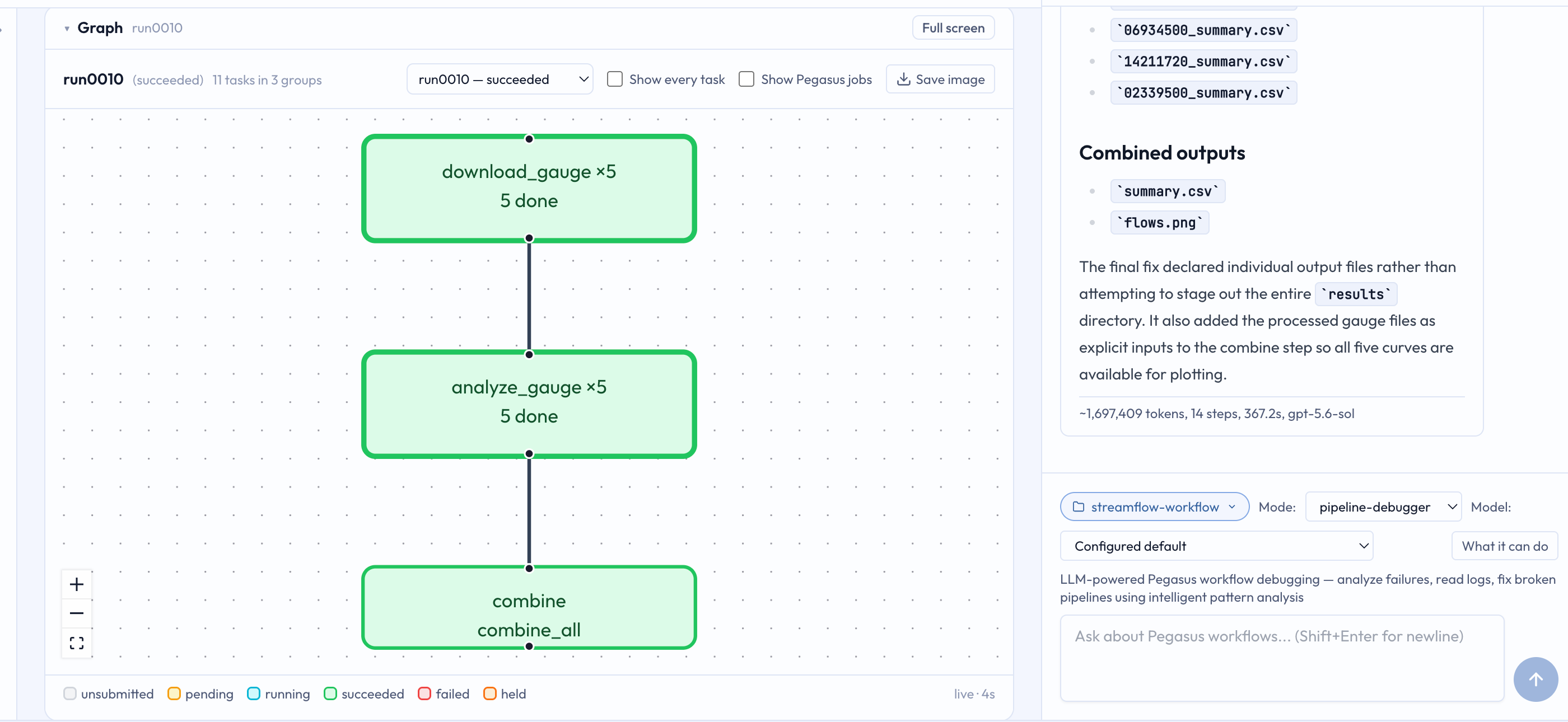}
\caption{Studio's live graph view of a completed run, driven by the same
event stream the agent reads, beside the chat in its workflow-debugger
mode. The workflow (five stream gauges fetched, analyzed, and combined)
was built from one chat prompt, and the chat transcript on the right is
the agent explaining the fix that made this run succeed after an earlier
failure. Node colors track live scheduler state (succeeded, running,
pending, failed, held).}
\label{fig:studio-dag}
\end{figure*}

\textbf{Monitor.} A monitoring service follows each run by reading the
Pegasus Stampede workflow database~\cite{gunter2011stampede,vahi2012monitoring}
and scheduler state, and emits a
structured JSONL event stream: job state transitions, hold reasons, stage-in
and stage-out progress, and resource usage. This service is required
because the state of a run is split between two stores that are built for
querying rather than for streaming: the workflow database knows which jobs
exist and what each one produced, while the scheduler knows which jobs are
queued, running, or held, and why. The service merges the two into one
time-ordered stream that the user interface renders and that the agent
reads as its window into execution. The agent therefore
reasons from structured workflow and scheduler state rather than scraping
terminal output.
Figure~\ref{fig:studio-dag} shows the human-facing rendering: a live
graph of a run, with each node colored by its current scheduler state and
each step labeled with the execution host it ran on. The monitor also flags a
step that has run at least ten minutes and five times longer than the
median successful duration of the same step in this run and the previous
five runs of the workflow. That check came from an annex test in which a
data-fetch step that usually takes seconds held a worker slot for forty
minutes while the run page showed no failure.
The graph collapses the fan-out of one step over many inputs into a
single node with a count, can expand to every task and to the auxiliary
Pegasus jobs, and can show the files each step reads and writes as nodes
of their own, with a preview of each file.

Figures~\ref{fig:studio-form} and~\ref{fig:studio-dag} show the same
streamflow workflow, a tutorial example built inside Studio. A single request asked the agent to fetch 30 days of discharge for
five USGS gauges, compute per-gauge statistics and a seven-day rolling average,
produce a combined table and plot, parameterize the gauge list and lookback
window, and tolerate individual fetch failures while failing the run if fewer
than two gauges returned data. The assistant scaffolded the workflow, ran it,
and worked through failed runs using the diagnose step until the execution in
Figure~\ref{fig:studio-dag} succeeded. In this case the failure information becomes part of the same lifecycle that Studio monitors and the
agent diagnoses.

\textbf{Diagnose.} On failure, Studio collects the analyzer's report
(\texttt{pegasus-analyzer}), the relevant job logs, and the hold reasons,
and hands them to the chat engine as structured context. The diagnostic
skill matches that context against a table of recurring failure modes, each
recorded as an error signature, its cause, and a fix. The modes fall into four families:
\begin{itemize}
\item \textbf{Missing credentials on execute nodes.} HTCondor starts each
  job in a clean environment, so an API key exported in the submit shell
  never reaches the job or its container, and the fetch step fails on the
  worker with an error that does not name the cause. The fix is to read the key when the
  workflow is generated and inject it into the job's environment.
\item \textbf{Catalog and container mismatches.} A container declared
  without the site that holds its image makes planning fail with a
  container-site mismatch, followed by an error that no physical file name
  can be selected for the image. A separate signature, an image that is
  ``not a valid squashfs image'', traces to a build that reported success
  but produced an empty container.
\item \textbf{Stage-out failures that hold a DAG.} When a wrapper exits
  before it writes its declared outputs, HTCondor cannot transfer them and
  places the job on hold. It reports the missing files, not the cause, and
  the workflow stops making progress. The skill directs the agent to the
  job's own error log, where Pegasus records the real failure.
\item \textbf{Site misconfiguration.} Planning fails when no site catalog
  defines the requested execution site, or when the generator registered
  its transformations for a different site than the one the workflow is
  planned for. Studio avoids the second case by passing the same site to
  both steps from the run panel's choice of where the workflow runs.
\end{itemize}
The list grows as new failures are traced to recurring causes.

\textbf{Fix and resubmit.} Proposed changes are ordinary edits to the
workflow project, made through the agent's guarded tools. Studio checks
each edited generator for syntax errors and for a parseable argparse
interface before the workflow is replanned, so an edit cannot silently
break the interface contract of Section~\ref{sec:knowledge}.

\begin{figure*}[t]
\centering
\includegraphics[width=0.95\textwidth]{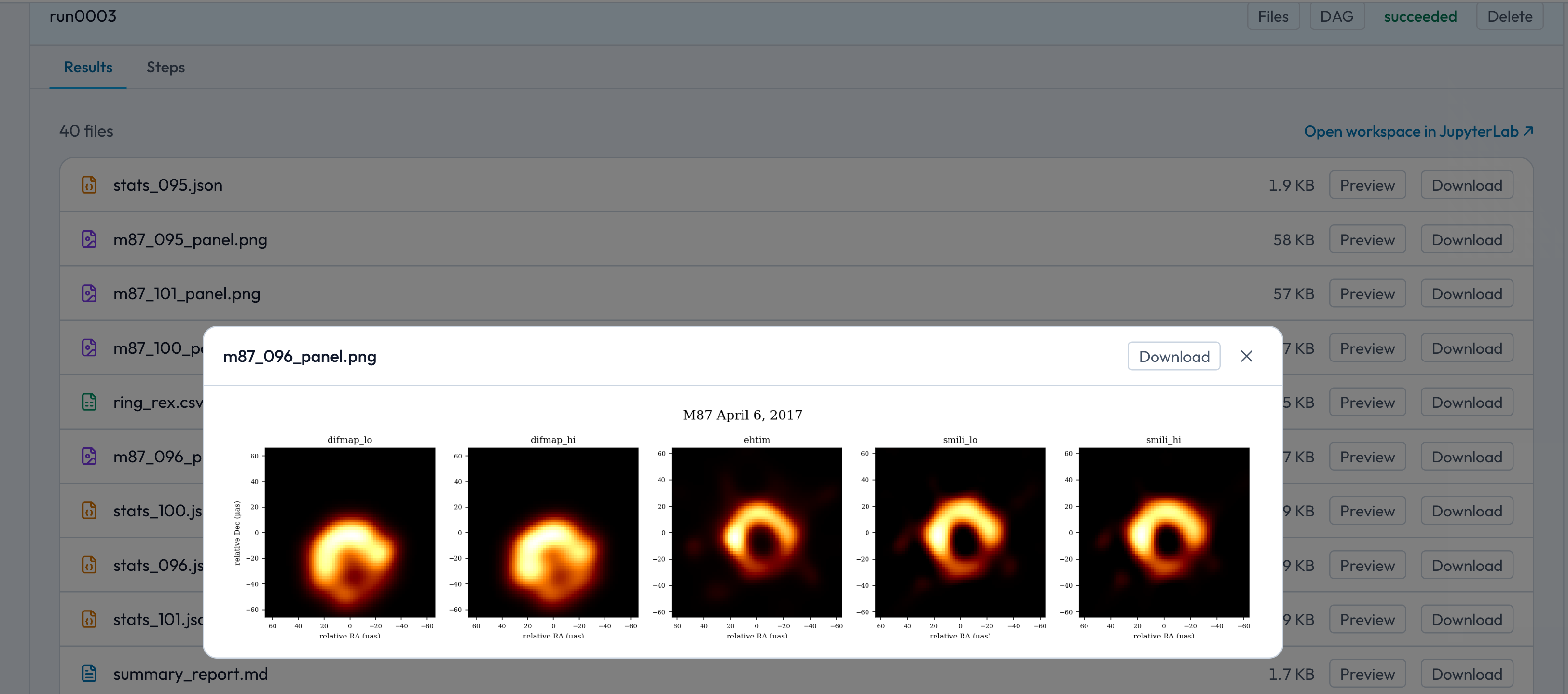}
\caption{The results stage for the EHT reconstruction's successful run inside
Studio. The run lists its 40 result files (per-day panel images, closure
statistics, the ring-measurement table, and the summary report), each
with a preview and a download, and a link that opens the run directory
in JupyterLab. The preview shows the April~6, 2017 reconstruction by all
three pipelines, one day's reconstructions produced on the gateway pool.}
\label{fig:studio-results}
\end{figure*}

\textbf{Results.} The lifecycle ends with the scientist examining the workflow's outputs, and Studio treats that as a lifecycle stage rather
than leaving the user to find files in a submit directory. Each run
lists its result files, and clicking a step in the graph shows the files
that step read and wrote. Common result formats open in place: tabular
files render as tables, images and PDFs inline, and long text files as a
partial preview. Every inline preview except raster images and PDFs is
served sandboxed, because workflow outputs such as HTML reports can carry
scripts. The same run
directory opens in JupyterLab for further analysis, which
is how the EEG and streamflow outputs in this article were inspected. The
agent reads the same files through its tools, so a question such as
``does the summary table have a row for every gauge'' is answered from
the result itself, not from the job log. For a scientist this is the
step that turns a successfully executed workflow into a result they can actually inspect and use. Figure~\ref{fig:studio-results} shows the stage for
the EHT reconstruction's run inside Studio (Section~\ref{sec:capstone}): the
run's result files, with one day's reconstructed images opened in place.

Studio is introspective in that it instruments its own assistance lifecycle.  For every failed run it keeps one record, which we call a \emph{failure
episode}. The episode follows what happens after the failure: whether the user asked the assistant about it, whether the user re-ran the same
workflow, and whether that re-run reached a terminal state. Studio stores
each of these steps with its time in the user's own database. At the point where the answer is meaningful,
after the re-run finishes, the user is asked one question: did the
assistant's suggestion fix it (Figure~\ref{fig:studio-feedback} in
Section~\ref{sec:trust})? The prompts are rate-limited server-side so
that they do not become a survey users learn to dismiss. This is the
per-episode measurement that the command-line setting lacked
(Section~\ref{sec:eval-evidence}). 
At the time of writing, no feedback data had yet been collected for evaluation.

\subsection{Why structured signals matter}

Grounding the agent in structured workflow and scheduler telemetry  has two effects. It
makes agent conclusions checkable, because every claim about a run traces
to an event or analyzer line a human can inspect. It also gives the diagnose, fix, and resubmit lifecycle a
concrete starting point:  the diagnose step starts from the actual failure
record rather than from a user's paraphrase of it. The CASA nowcast case
study (Section~\ref{sec:casa}) shows this process identifying three real defects
that terminal output alone had obscured.

In the current implementation, \texttt{workflow-monitor} is a read-only tool alongside Pegasus. Its
integration into Pegasus itself through the MonitorD plugin system is in
progress and is discussed further in (Section~\ref{sec:monitor}).

%% file: sections/10-trust.tex
\section{The Trust Envelope: Making Agents Accountable to CI}
\label{sec:trust}

The command-line construction setting used the operator's shell,
credentials, and model configuration~\cite{pegasus_method_companion}. A
shared facility needs a different operational boundary: it must be able to attribute an
agent's jobs and resource use to an authenticated user and to limit the
agent to that user's privileges. This requirement is particularly important
for AI agents because they can initiate actions that the user has not individually
specified. Studio therefore runs agent actions under the user's
identity and combines that identity with explicit controls over which
state-changing actions may proceed.

\subsection{Identity: the agent is tied to the user}

The chain runs CILogon to Authentik to JupyterHub to a per-user pod. At
spawn, the platform mints an HTCondor IDTOKEN for the authenticated user and
mounts it, together with the user's NFS home, into the pod. Every process in
the pod, including every AI agent, runs under the user's UID with the user's
token. A job submitted by an agent is, to the scheduler and to the audit
trail, a job submitted by that user. There is no service account, no shared
credential, and no privilege the agent holds that the user does not.

The same rule holds in the Open OnDemand deployment described in
Sec.~\ref{sec:tiers}, where the portal authenticates the user, the
session runs under the user's POSIX account on a compute node, and
workflow jobs enter the cluster's Slurm queue under that account. There
Studio keeps its own home directory and never writes into the user's
real home, because a cluster home already holds the user's own
configuration of the same agent tools.

The rule also extends to resources the user brings. A user whose
workflows need more than the shared pool can lease nodes from Slurm
clusters on which they hold an allocation, including campus clusters or ACCESS
resources. This is accomplished using HTCondor's Annex mechanism. Studio generates a
self-contained job that the user submits to the remote cluster's Slurm
queue, and the nodes it acquires join the pool as an annex named after
the user. The run form lets the user send a workflow's jobs only to that
Annex (Figure~\ref{fig:studio-form}), and only that user's jobs may run on
it, so the allocation
remains associated with the user who owns it, even when an agent submits work on that user's behalf. Idle Annex nodes shut themselves down. The first live
Annex test, on SDSC Expanse~\cite{strande2021expanse}, ran the EHT  workflow of
Section~\ref{sec:capstone} to completion with all 31 of its compute steps
on the user's annex nodes.

\subsection{Authorization: approval posture}

Identity links an agent's actions to a user. Authorization decides
which of those actions the agent may take on its own. Every coding agent
in Studio has a built-in permission mechanism: before the agent writes a
file, runs a command, or submits a workflow, it can be configured to
present the proposed action to the user and wait for a yes or no. Studio
configures every agent with an approval policy for state-changing actions, with two
refinements learned while exercising the deployment. Edits inside the user's workflow and
data directories proceed without a prompt, while writes anywhere else in the
home directory always ask, because an unprompted write to a shell profile or
to the agent's own configuration could alter the agent's permission configuration.
Workflow submission always asks, even when the user has turned off
confirmation for routine shell commands. We describe
the resulting autonomy on a three-level scale. At \emph{assist}, the agent
only answers questions and proposes changes, and the user applies them. At
\emph{approve}, the agent applies changes and runs commands itself, but
state-changing actions outside the user's workflow and data directories,
and every workflow submission,  require the user's confirmation. At
\emph{autopilot}, the agent acts without confirmation. The deployed
default is \emph{approve}: the agent may operate within the user's project workspace, while higher-risk
writes and workflow submission remain approval-gated.  Read-only
actions (inspecting events, reading logs, querying documentation) need no
approval and flow freely, which keeps the diagnose step fast without
widening the write path beyond the user's own project files.
Section~\ref{sec:discussion} explains why we have not moved any action
class to autopilot.

\subsection{Model governance: provider-neutral access}

Institutions differ in which model providers are acceptable, for reasons of
cost, procurement, and data governance. Studio treats the model endpoint as
configuration: commercial APIs, campus or national research gateways, or
local models can be changed per user, and a user may bring their own
provider key. Prompts that include research data can therefore be kept on
infrastructure the institution controls. No Studio capability assumes a
particular provider, and the platform remains usable without a model provider.

The public deployment routes model requests through a platform model gateway that
exposes one virtual model name to every agent and resolves it in tiers.
The first tier is a set of open-weight models hosted on research clouds
(at the time of writing, Qwen~\cite{qwen2025qwen3} models served by the National Research
Platform~\cite{nrp} and by Jetstream2~\cite{hancock2021jetstream2}), load-balanced by remaining request budget
and latency. A commercial endpoint is the second tier, tried only when
all first-tier backends have failed for a request. Each backend carries
its own request-rate limit and circuit breaker. In front of the gateway
sits a prompt guard that scans every input for prompt-injection patterns
and redacts bearer tokens before they reach a model, and scans outputs
before they return. The practical effect is that a scientist's prompts
go to a research-cloud model by default and reach a commercial provider
only as a fallback, while an institution can change that policy in
one place.

\subsection{Observability and consent}

Accountability also requires that the facility can see what its agents
cost and do. The chat engine in every container emits OpenTelemetry~\cite{opentelemetry}
signals to a collector in the platform. Metrics, which record aggregate
activity (tokens by type, tool-call durations, sessions and messages,
model and provider, per user), always flow and feed a dashboard that
shows usage and cost by user, which agent role handled each message, and
the health of every gateway backend. Traces, which carry the prompt text
and tool arguments, flow only for users who have opted in. 

The guided tour asks the user
to make this choice with no preselected answer, and the choice can later
be changed in settings. If the setting cannot be read, Studio defaults to
no consent. Consent also travels with the data as an attribute on every
span, so the collector strips content from any trace whose resource does
not assert consent, even if a running container still has a configuration
written before the user changed the setting.
Traces are stored in a trace
store inside the cluster and are not exposed publicly. This constitutes the
evidence base upon which the autonomy discussion in Section~\ref{sec:discussion}
depends on.

Table~\ref{tab:collected} lists everything Studio records about a
user's interaction with the assistant, and under what condition. A
collection notice at the end of the guided tour states the same list in
plain language.

\begin{figure}[t]
\centering
\includegraphics[width=\columnwidth]{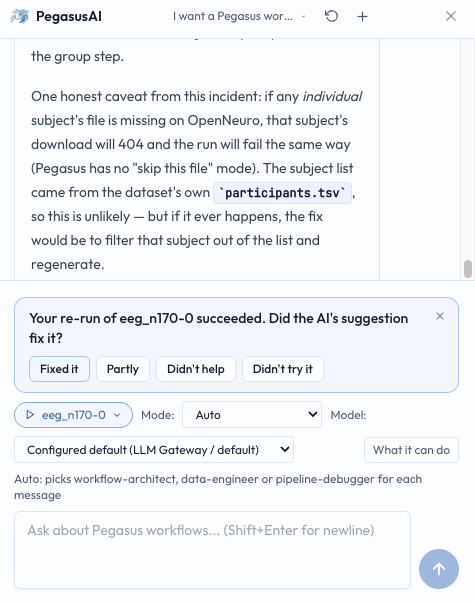}
\caption{The help card. After a failed workflow has been re-run and the
re-run has finished, Studio asks one question at the point where the
answer is meaningful: did the assistant's suggestion fix it? The four
answers are one click each, and the card is rate-limited so that it
does not become a survey users learn to dismiss. Here the EEG workflow
of Section~\ref{sec:eeg} succeeded on its fourth run after the
assistant diagnosed the earlier failure. The chat is in its default
mode, which routes each message to a specialist agent role.}
\label{fig:studio-feedback}
\end{figure}

The tour continues
to request an explicit answer about whether chat text may be recorded until
the user answers either way. Optional profile questions are handled
separately. 
Three kinds of
record carry no user text and are always kept: the usage metrics mentioned above,
the per-failure episodes of Section~\ref{sec:lifecycle} together with
the user's one-click answer to the help card of
Figure~\ref{fig:studio-feedback}, and the ratings a user gives
individual replies (a thumbs up or down, or a tag when the user stops a
reply). The fourth kind, the text of prompts and responses and the
arguments of tool calls, is the conversation itself and is kept only
for users who opted in. The purpose is stated to the user as well: the
episodes and ratings are how the facility will learn which classes of
agent action 
tend to succeed without human correction, while the conversation
text helps it learn why the other actions do not.

\begin{table}[t]
\centering
\caption{What Studio records about a user's interaction with the
assistant. The first three rows carry no user text. The last row is the
conversation itself and is recorded only with the user's opt-in.}
\label{tab:collected}
\small
\begin{tabular}{@{}>{\raggedright\arraybackslash}p{0.40\linewidth}>{\raggedright\arraybackslash}p{0.20\linewidth}>{\raggedright\arraybackslash}p{0.30\linewidth}@{}}
\toprule
Record & Condition & Kept where \\
\midrule
Usage metrics: tokens by type, tool-call durations, sessions and
messages, model and provider, retries, per user & always &
platform metrics store and the facility dashboard \\
\addlinespace
Failure episodes: a run failed, whether the assistant was asked, whether
a re-run followed and how it ended, and the user's answer to the help
card (fixed it, partly, did not help, did not try it) & always, one
question per episode & the user's own database, read by administrators from the shared home \\
\addlinespace
Reply ratings (thumbs up or down) and give-up tags, and the optional
profile answers from the tour & when the user clicks or answers &
the user's own database, rating counts also as platform metrics \\
\addlinespace
Prompt and response text, tool-call arguments (including shell
commands) & only with opt-in, revocable in settings & in-cluster trace
store, the collector strips content otherwise \\
\bottomrule
\end{tabular}
\end{table}

None of these mechanisms require changes to Pegasus or HTCondor. Studio uses
the existing token mechanism, scheduler authentication, and WMS provenance
records as provided by those systems. This is important for the trust
argument: a resource provider can evaluate agent-driven execution in terms of
the same identity, authorization, and audit mechanisms it already uses for
human-submitted workflows.

%% file: sections/11-evaluation-studio.tex
\section{Evaluation and Case Studies}
\label{sec:evaluation}
The evaluation examines whether Studio's mechanisms function during real workflow
development and execution. 

The evidence here comes from case
studies and deployment tests and is observational rather than a
controlled comparison.

\subsection{Research questions}
\label{sec:rqs}
\begin{itemize}
\item \textbf{RQ1 (Telemetry grounding).} Does structured workflow and
scheduler state let the agent find and repair real execution failures?
\item \textbf{RQ2 (Interface legibility).} Can an agent-written workflow
become operable from the Studio interface without hand integration?
\item \textbf{RQ3 (Accountability).} Do agent actions, including repairs
and resubmissions, remain attributable to the user and subject to the configured approval
policy?
\item \textbf{RQ4 (Generality across deployments and agents).} Does the same application image operate across both 
deployment models, and does the projected knowledge work with more than one agent and
model stack?
\end{itemize}

With the supporting evidence in the subsections below, we can summarize the findings for each research question as follows:

For \textbf{RQ1}, in the cases examined, the evidence supports telemetry-grounded diagnosis and repair.  Telemetry exposed three
stacked defects in the CASA nowcast port, which then completed all 86 jobs
(Section~\ref{sec:casa}). In the EHT reconstruction, the agent went from two failed
runs to a zero-failure run in under an hour (Section~\ref{sec:capstone}).

For \textbf{RQ2}, Studio derived the \texttt{gw} workflow's parameter
form from the agent's generator, the first launch exposed a defect at that
boundary, and the corrected workflow ran 350 of 350 jobs from the same form
(Section~\ref{sec:gwcase}).


For \textbf{RQ3}, in the EHT reconstruction, the rebuild and resubmission ran
under the user's scheduler identity and remained subject to the user's configured approval policy
(Section~\ref{sec:capstone}).


For \textbf{RQ4}, the same application image ran as a JupyterHub gateway and as an Open
OnDemand application on a test cluster, and the gateway ran the EHT
workflow on a user's annex at SDSC Expanse. The projected knowledge was also exercised with more than one agent and
model stack (Section~\ref{sec:generality}).


\subsection{Scope of the evidence}
\label{sec:eval-evidence}

The evaluation focuses on cases in which a Studio mechanism and an observed
outcome are directly linked. In the CASA nowcast port
(Section~\ref{sec:casa}), workflow and scheduler telemetry exposed three
stacked execution problems and the repair cycle ended in a verified 86-job
run. In the Gravitational-wave workflow (Section~\ref{sec:gwcase}), Studio
constructed its parameter form from an agent-written generator. The first
launch exposed an interface defect at that boundary, and the corrected
workflow then ran to completion from the same form. 
In the EEG analysis (Section~\ref{sec:eeg}), the projected knowledge base was exercised with a different agent and model stack, and the resulting workflow recovered from an execution-environment failure to complete all 127 jobs successfully.
In the EHT reconstruction
(Section~\ref{sec:capstone}), two gateway runs failed at stage-in. The debug
path supplied the analyzer report and job logs to the workflow-debugger agent,
which identified missing container images, repaired their unprivileged builds
under user approval, and resubmitted to a zero-failure run. These cases show
the mechanisms operating as designed. They do not quantify how much effort the
mechanisms save.

\subsection{CASA nowcast: telemetry-grounded diagnosis}
\label{sec:casa}

CASA (Collaborative Adaptive Sensing of the Atmosphere) operates a network of
short-range weather radars~\cite{mclaughlin2009casa} whose data feed operational nowcasting products. A
legacy CASA nowcasting workflow, originally written for an earlier Pegasus
release, was modernized for Pegasus 5 inside Studio and verified end to end:
all 86 jobs succeeded. The diagnose step surfaced three stacked defects that
were not apparent from terminal output alone. Unprivileged pods cannot run Docker, so execution
had to move to Apptainer. A leftover working-directory catalog silently
shadowed the embedded catalogs, which was revealed by a malformed URL in the planned
transfer list. A lax worker-package setting then let the host's newer GLIBC
kickstart run inside an older container, breaking jobs at startup. Each failure
was identified from workflow or scheduler state, fixed, and re-run through the agent-assisted 
diagnose, fix, and resubmit loop of Section~\ref{sec:lifecycle}. The resulting port was submitted as an upstream
pull request.

Two sibling CASA workflows were modernized in the same sessions and verified in 
the same way (Wind at 23 of 23 jobs and Hail at 14 of 14). 

The CASA cases also show that legacy Pegasus workflows can be brought into the Studio lifecycle without changes to Pegasus or HTCondor.

\subsection{Gravitational waves: exercising the interface contract}
\label{sec:gwcase}

The \texttt{GW} workflow~\cite{pegasusai-workflows} reanalyzes GWTC-1 binary black
hole events~\cite{abbott2019gwtc1} from GWOSC open strain
data~\cite{abbott2021gwosc}. The workflow was built entirely inside Studio, using Claude Code
with Claude Opus~5. 
 
The 
question is whether an agent-written workflow can become directly operable
through the Studio interface.

Studio derived the workflow's parameter form from the generator's
\texttt{argparse} interface without executing the code. The first launch from
that form exposed a real boundary defect: the form submitted a
comma-separated list where the command line expected separate tokens. The
agent changed the interface to accept both representations, after which the
same form launched the run of record. It completed 350 of 350 jobs (287 tasks)
with zero failures in 54 minutes 18 seconds on the Studio pool, and its scientific
checks passed.  The case demonstrates the structural-legibility
contract of Section~\ref{sec:knowledge}: an interface written by the agent was
introspected by Studio, rendered to the user, exercised, corrected at the
boundary, and then used to launch the completed workflow.

\subsection{EEG analysis: exercising knowledge projection across an agent and model stack}
\label{sec:eeg}

The EEG workflow provides an end-to-end test of Studio's knowledge projection
with a different agent and model stack. OpenCode, using the open-weight
Kimi K2.6 model~\cite{kimi2025k2}, operated entirely inside Studio from a
single request for a group-level N170 face-perception analysis over OpenNeuro
dataset ds007096~\cite{markiewicz2021openneuro}. Working from the projected
knowledge base, the agent invoked the scaffold skill and produced a
per-subject fan-out followed by a group-level fan-in.

The same session exercised the debugging path. Container wrapping failed on
the heterogeneous gateway workers, and the agent recovered by replacing the
container with a portable bundle of the required Python dependencies staged
as a workflow input. The final run over 24 subjects completed all 127 jobs
(97 tasks) in 5 minutes 40 seconds with zero failures and zero retries.

\subsection{EHT reconstruction: telemetry-grounded debugging and accountable execution}
\label{sec:capstone}

The EHT reconstruction provides an end-to-end test of Studio because a
scientifically substantive workflow encountered real infrastructure failures
while running through the gateway. 
The case exercises several Studio mechanisms
together: exposing failure state to the agent, repairing the workflow under the
user's authority, resubmitting it through the same execution path, and returning
the resulting scientific outputs through the workflow lifecycle.

The reconstructed workflow~\cite{eht-workflow}, which builds on an earlier Pegasus-based reproduction of the EHT M87 imaging pipelines~\cite{DBLP:journals/cse/PatelRCKLTVBDT22}, contains 31 scientific tasks across the DIFMAP, eht-imaging, and SMILI imaging pipelines~\cite{ehtIV2019} plus rendering, closure statistics,
ring measurement, and aggregation. Four containers reproduce the required
software environments, three of them from the 2019 software era. The workflow
was built and executed inside Studio on the gateway's HTCondor pool, where its
container images had to be built without root privileges in the user's pod.

The first two gateway runs failed at stage-in because three of the four images
had not been produced successfully by the unprivileged container builds.

The user invoked
Studio's debug path. The workflow-debugger agent received the analyzer report
and job logs, identified the missing images, the repeated failure, and
explained the diagnosis. With the user's approval, the agent rebuilt the images. Those
builds exposed three problems specific to unprivileged reconstruction of the
old software stack: an \texttt{apt} sandbox-user failure inside a user
namespace, an archived Debian package URL that returned 404, and a
tarball removed before it was needed. The agent repaired the definitions,
verified each image 
by running an import test for its software inside the container,
regenerated
the workflow, and resubmitted it.

\begin{figure*}[t]
\centering
\includegraphics[width=0.95\textwidth]{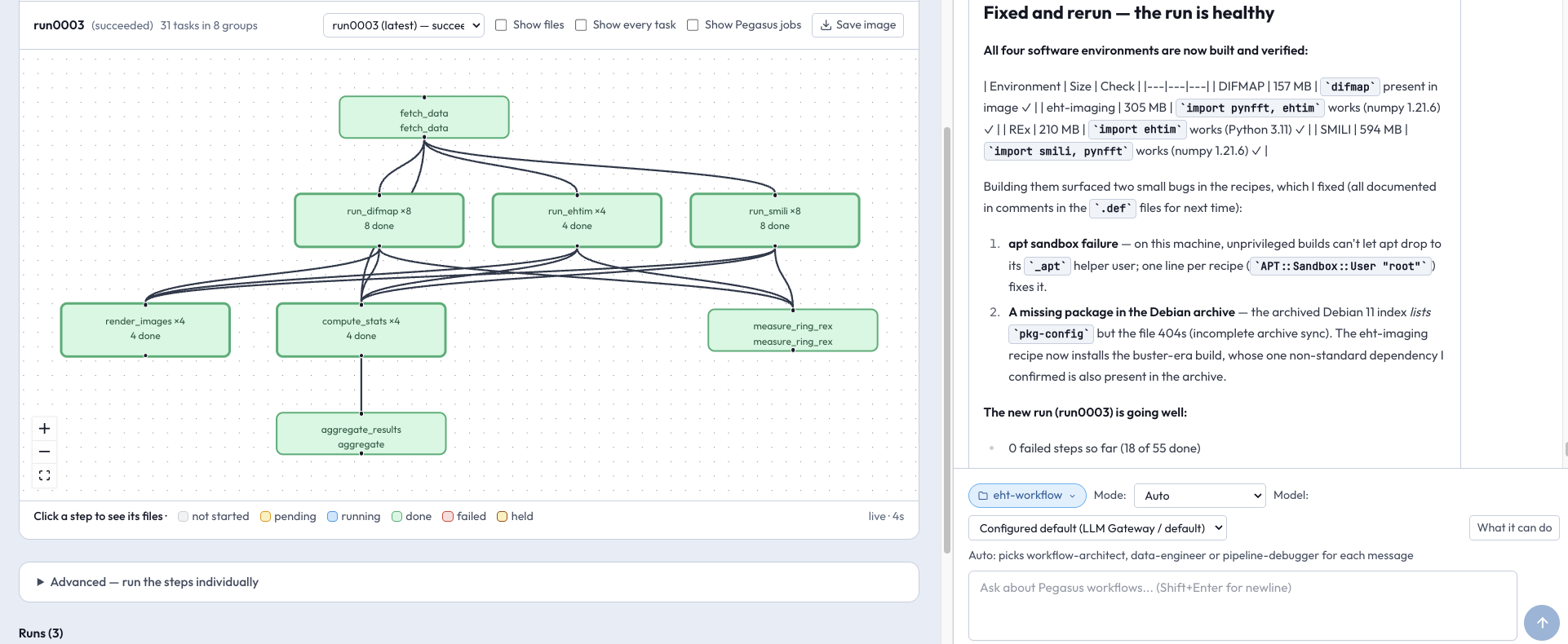}
\caption{The EHT reconstruction inside Studio. Left: the graph of the successful
third gateway run, with the per-day fan-out of the three imaging pipelines
collapsed into one node per pipeline. Right: the \textit{workflow-debugger} agent's
summary after rebuilding and verifying the four container images. The same
workflow and scheduler records that drive the graph supplied the diagnostic
context to the agent.}
\label{fig:studio-eht}
\end{figure*}

The third gateway run completed 55 of 55 jobs (31 tasks) with zero failures and
zero retries in 8 minutes 34 seconds, less than an hour after the first failure
and with two user prompts in the chat. Figure~\ref{fig:studio-eht} shows the
successful graph and the \textit{workflow-debugger's} closing summary. The same episode also exercises Studio's accountability mechanisms: the agent ran in the authenticated user's Studio
container, jobs carried that user's scheduler identity, and the rebuild and
resubmission path remained subject to the approval process of
Section~\ref{sec:trust}. The failed runs, diagnosis, edits, and successful run
remain visible in the workflow and scheduler records that supplied the agent's
telemetry.

The successful run produced the reconstructed M87 images, closure statistics,
and ring measurements shown through Studio's results interface in
Fig.~\ref{fig:studio-results}. 

The EHT case provides systems evidence: telemetry from the failed executions reached the agent, the repair occurred under the
user's identity and approval, and the corrected workflow completed on the
shared gateway pool.

\subsection{Generality across deployments and agents}
\label{sec:generality}

\textbf{Portability across deployment environments.} Studio uses the same application image in two
deployment models: a JupyterHub/Kubernetes gateway on Jetstream2 and an Open
OnDemand application for HPC systems. The gateway has executed workflows on
its shared HTCondor pool and, through a user's annex on SDSC Expanse, ran the
EHT reconstruction with all 31 compute steps on the user's own allocation
(Section~\ref{sec:trust}). The OOD deployment has been validated on a test cluster
(Section~\ref{sec:tiers}).
 
These observations
demonstrate that the Studio architecture operates across the environments
actually tested, not across workflow systems in general.

\textbf{Knowledge projection across agent/model stacks.} The canonical knowledge base of
Section~\ref{sec:knowledge} is projected into five agent-native
configurations. The \texttt{GW} case above used the Claude Code projection. 
The \texttt{EEG} exercised the same projection mechanism with a different agent and model stack: OpenCode driving
Kimi K2.6~\cite{kimi2025k2}.
Together, these cases show that the projected
knowledge and lifecycle operate with more than one agent and model stack.
They do not establish agent independence or compare model capability.

%% file: sections/12-discussion.tex
\section{Discussion and Lessons Learned from Deployment}
\label{sec:discussion}

\subsection{Enhancing vetted infrastructure rather than replacing it}
\label{sec:enhance}

A common direction in AI for science is to recast scientific computation as
an agentic workflow, in which a language-model agent plans the steps, calls
tools, and drives execution itself~\cite{boiko2023coscientist,bran2024chemcrow,ma2025llmhpc}.
That design is simplest when the agent controls the resources it uses.
Research cyberinfrastructure rarely offers that. Compute, storage, and data
services belong to facilities and projects that set their own policies for
identity, allocation, scheduling, data movement, and accounting. An agent
that orchestrates these resources directly has to reproduce each of those
controls or work around them. It also has to recover from the partial
failures of distributed execution, such as held jobs, failed transfers, and
lost nodes, which workflow systems such as Pegasus~\cite{deelman2015pegasus}
were built to handle.

Studio takes the other path. It leaves execution to a vetted stack and
applies AI where that stack is hard to use. Pegasus plans the workflow,
retries failed jobs, and records provenance. HTCondor and the site's batch
scheduler place jobs and account for them. JupyterHub, Open OnDemand, and
CILogon establish who the user is. The agent writes and repairs the
workflows that this stack runs, and it reads the stack's own state to
explain failures. Because it acts under the user's identity, it never holds
access to facility resources that the user does not already have. The capabilities and controls that
facilities already trust are preserved, and AI lowers the cost of
using them. We expect the same pattern to apply wherever the resources an
agent would act on are owned and governed by a third party.

\subsection{What generalizes beyond Pegasus}
The systems mechanisms
in Studio are expressed in terms of abstractions that are not unique to
Pegasus.  Knowledge projection
requires a canonical body of system and site knowledge plus adapters to the
configuration interfaces of supported agents. Telemetry grounding requires
queryable workflow and scheduler state that can be turned into structured
execution signals. The trust envelope requires an existing user identity,
authorization boundary, and audit trail under which agent actions can run.
Pegasus, HTCondor, JupyterHub, and Open OnDemand provide those mechanisms in
Studio. The architecture is designed around these abstractions rather than their particular APIs,
although it has so far been implemented only with them
(Section~\ref{sec:threats}).

For the scientist, the resulting division of labor is similar across systems:
the agent can construct and operate workflow machinery, while the human remains
responsible for scientific intent, review, and validation. For facilities, the
important shift is that workflow and site expertise becomes maintained,
versioned infrastructure rather than context configured separately for each
user and agent.

\subsection{Practical lessons from deployment}

Four lessons from operating the gateway are not specific to Pegasus.

\textbf{Reverse-proxy prefixes break assumptions.} Per-user URL prefixes under
JupyterHub exposed assumptions about build-time base paths, asset paths, and
WebSocket origins. Applications intended for gateway deployment need these
constraints accounted for in the design rather than discovered after
integration.

\textbf{Streaming requires infrastructure support.} Default ingress request
timeouts silently reset agent and chat streams. Routes carrying server-sent
events or long-running tool calls need an explicit timeout policy. The failure
mode is largely invisible in local development.

\textbf{Agents inherit the execution environment's sharp edges.} Clean job
environments, nested container constraints, per-user token trust domains,
unprivileged pods, and old submit-host software are infrastructure properties,
not AI-specific ones. The knowledge layer of Section~\ref{sec:knowledge} is
where these lessons accumulate so that later agent sessions do not have to
rediscover them.

\textbf{Autonomy must follow evidence.} Studio currently uses the
\emph{approve} posture described in Sec.~\ref{sec:trust}, in which higher-risk writes and every workflow submission remain approval-gated. Moving a class of actions,
such as resubmitting after a repair, toward unattended execution would require
evidence that the action can be carried out reliably without human correction.
The failure episodes and feedback described in Secs.~\ref{sec:lifecycle}
and~\ref{sec:trust} are intended to provide that evidence. Read-only actions
and edits inside the user's own workflow and data directories already
proceed without approval. Writes elsewhere and every submission do not.

\subsection{Demand signal}
\label{sec:monitor}
Requests for a version that institutions could deploy on their own clusters
led to the Open OnDemand deployment (Sec.~\ref{sec:tiers}), which has been
validated on a test cluster but not yet deployed at a production site.
Separately, the monitoring tool is being integrated into Pegasus through its plugin system. 

These are signs of interest, and they suggest that deployment portability and tighter monitoring integration are useful directions. They are not evidence that any institution has put Studio into production use, and we do not claim that.

%% file: sections/13-threats-studio.tex
\section{Limitations of the System and Evaluation}
\label{sec:threats}

\textbf{Self-development and limited users.} Studio has so far been used
mainly by its own developers and by collaborators who are already familiar with Pegasus.
Thus, the case studies show that the mechanisms operate as designed, not
that they reduce effort for a representative user population. The failure
episodes and help-card answers of Sections~\ref{sec:lifecycle}
and~\ref{sec:trust} are designed to collect evidence about agent assistance as the user population broadens.

\textbf{No quantitative comparison of debugging effort.} 
The case studies show that Studio exposes structured state and supports
diagnosis and repair, but they do not establish that it reduces debugging
effort relative to other ways of operating the same workflows.

A controlled comparison remains future work.

\textbf{Case selection.} 
The authors selected all of the case studies. The
workflows span different scientific domains, software stacks, and execution
patterns, but they were not sampled systematically from a broader population
of scientific workflows. The evaluation therefore demonstrates the Studio
mechanisms across the cases examined, but does not establish how representative
those cases are of scientific workflow development more broadly.

\textbf{Generality is partly a design claim.} Knowledge projection,
telemetry grounding, and the trust envelope are built from abstractions that
exist in other workflow systems and gateways. They have been implemented and
tested only with Pegasus, HTCondor, JupyterHub, and Open OnDemand, and the
Open OnDemand deployment has been validated only on a test cluster.

\textbf{Current implementation limits.} The application container is
single-user, some state is SQLite-backed, AI tools are installed per
container, and some model credentials are stored locally. These are limits
of the current implementation rather than requirements of the architecture.

\textbf{Agent and model drift.} Case-study outcomes depend on the agent,
model, and version used, and these change quickly. We report the agent and
model for each case. The cross-agent evidence shows interoperability where
tested. It is not a benchmark of model capability.

%% file: sections/14-artifact-studio.tex
\section{Artifact Availability}
\label{sec:artifact}
PegasusAI Studio is open source~\cite{pegasusai-studio}, as is the Open
OnDemand application~\cite{pegasusai-ood}. 
The Studio release also provides
links to the public gateway and project site. 
The deployment repository contains
site configuration and sealed secrets and is therefore not released as-is.
Reviewable manifests can be provided separately.

The canonical PegasusAI knowledge base, including skills, agent roles,
templates, and documentation servers, is open source in the PegasusAI plugin
marketplace~\cite{pegasusai-plugin}. All four case-study workflows are public
repositories. The CASA nowcast port (\texttt{casa-nowcast-workflow}), the
gravitational-wave reanalysis (\texttt{gw-workflow}), and the EEG analysis
(\texttt{eeg-workflow}) are in the \texttt{pegasus-isi} GitHub
organization~\cite{pegasusai-workflows}. The EHT reconstruction is
published separately~\cite{eht-workflow}. 

%% file: sections/15-conclusion-studio.tex
\section{Conclusion}
\label{sec:conclusion}
An AI agent that writes a workflow on a laptop has solved only part of the workflow execution 
problem. Shared CI also requires the agent to know the
system's conventions, to monitor the state of distributed executions, and to act
under an identity the facility can accept and audit. 
We presented PegasusAI Studio, which makes CI legible to agents through knowledge projection, interface and site introspection, and telemetry grounding. In addition, it makes agents accountable through federated identity, per-user scheduler credentials, approval controls for state-changing actions, and observability. 
Rather than reimplementing execution infrastructure around the AI agent,
Studio builds these capabilities on mature, production-proven CI systems.
Its design follows one rule. Mature execution systems operate the workflow, so an unmodified Pegasus and HTCondor stack handles scheduling, data movement, recovery, provenance, and resource management. Mature
gateway systems provide identity and multi-user isolation, so JupyterHub and Open OnDemand authenticate each user and give each one a separate
environment. Studio adds only the integration layer that is specific to
agents: projected knowledge, structured execution state, and controls on
what an agent may do.

The case studies show the mechanisms operating on real workflow development and execution, from legacy CASA ports to the EHT M87 reconstruction. They do not yet quantify how much effort the mechanisms save.
Consequently, our future work will include a more rigorous evaluation framework, which will analyze 

controlled comparisons of diagnosis and repair cycles, a broader population
of external users, and measured evidence for which classes of agent action
can safely move from approval to greater autonomy. Integrating the
monitoring service into Pegasus through its plugin system is the main engineering step in progress.

Studio is therefore best viewed as an operating layer between AI agents and
existing CI: maintain system and site knowledge as
versioned infrastructure and project it consistently to supported agents,
give agents the structured state the workflow and scheduling systems already
record, run every agent action under the user's own identity and let the
workflow and scheduling systems provide execution, recovery, and provenance.

%% file: refs.bib
@article{DBLP:journals/cse/PatelRCKLTVBDT22,
  author       = {Ria Patel and
                  Brandan Roachell and
                  Silvina Ca{\'{\i}}no{-}Lores and
                  Ross Ketron and
                  Jacob Leonard and
                  Nigel Tan and
                  Karan Vahi and
                  Duncan A. Brown and
                  Ewa Deelman and
                  Michela Taufer},
  title        = {Reproducibility of the First Image of a Black Hole in the Galaxy {M87}
                  From the Event Horizon Telescope Collaboration},
  journal      = {Comput. Sci. Eng.},
  volume       = {24},
  number       = {5},
  pages        = {42--52},
  year         = {2022},
  url          = {https://doi.org/10.1109/MCSE.2023.3241105},
  doi          = {10.1109/MCSE.2023.3241105},
  bibsource    = {dblp computer science bibliography, https://dblp.org}
}

@inproceedings{DBLP:conf/ucc/RoaROVMGKGHRLVL23,
  author       = {Camila Roa and
                  Mats Rynge and
                  Paula Olaya and
                  Karan Vahi and
                  Todd Miller and
                  James Griffioen and
                  Shelley Knuth and
                  John Goodhue and
                  D{\'{a}}vid Hud{\'{a}}k and
                  Alana Romanella and
                  Ricardo M. Llamas and
                  Rodrigo Vargas and
                  Miron Livny and
                  Ewa Deelman and
                  Michela Taufer},
  title        = {End-to-end Integration of Scientific Workflows on Distributed Cyberinfrastructures:
                  Challenges and Lessons Learned with an Earth Science Application},
  booktitle    = {Proceedings of the {IEEE/ACM} 16th International Conference on Utility
                  and Cloud Computing, {UCC} 2023, Taormina (Messina), Italy, December
                  4-7, 2023},
  pages        = {17},
  publisher    = {{ACM}},
  year         = {2023},
  url          = {https://doi.org/10.1145/3603166.3632142},
  doi          = {10.1145/3603166.3632142},
  bibsource    = {dblp computer science bibliography, https://dblp.org}
}

@article{deelman2015pegasus,
  author  = {Deelman, Ewa and Vahi, Karan and Juve, Gideon and Rynge, Mats
             and Callaghan, Scott and Maechling, Philip J. and Mayani, Rajiv
             and Chen, Weiwei and Ferreira da Silva, Rafael and Livny, Miron
             and Wenger, Kent},
  title   = {Pegasus, a workflow management system for science automation},
  journal = {Future Generation Computer Systems},
  volume  = {46},
  pages   = {17--35},
  year    = {2015},
  doi     = {10.1016/j.future.2014.10.008}
}

@article{thain2005condor,
  author  = {Thain, Douglas and Tannenbaum, Todd and Livny, Miron},
  title   = {Distributed computing in practice: the {Condor} experience},
  journal = {Concurrency and Computation: Practice and Experience},
  volume  = {17},
  number  = {2--4},
  pages   = {323--356},
  year    = {2005},
  doi     = {10.1002/cpe.938}
}

@article{fajardo2015htcondor,
  author  = {Fajardo, E. M. and Dost, J. M. and Holzman, B. and
             Tannenbaum, T. and Letts, J. and Tiradani, A. and
             Bockelman, B. and Frey, J. and Mason, D.},
  title   = {How Much Higher Can {HTCondor} Fly?},
  journal = {Journal of Physics: Conference Series},
  volume  = {664},
  number  = {6},
  pages   = {062014},
  year    = {2015},
  doi     = {10.1088/1742-6596/664/6/062014}
}

@article{ditommaso2017nextflow,
  author  = {Di Tommaso, Paolo and Chatzou, Maria and Floden, Evan W. and
             Barja, Pablo Prieto and Palumbo, Emilio and Notredame, Cedric},
  title   = {Nextflow enables reproducible computational workflows},
  journal = {Nature Biotechnology},
  volume  = {35},
  number  = {4},
  pages   = {316--319},
  year    = {2017},
  doi     = {10.1038/nbt.3820}
}

@article{molder2021snakemake,
  author  = {M{\"o}lder, Felix and Jablonski, Kim Philipp and Letcher, Brice
             and Hall, Michael B. and Tomkins-Tinch, Christopher H. and
             Sochat, Vanessa and Forster, Jan and Lee, Soohyun and
             Twardziok, Sven O. and Kanitz, Alexander and Wilm, Andreas and
             Holtgrewe, Manuel and Rahmann, Sven and Nahnsen, Sven and
             K{\"o}ster, Johannes},
  title   = {Sustainable data analysis with {Snakemake}},
  journal = {F1000Research},
  volume  = {10},
  pages   = {33},
  year    = {2021},
  doi     = {10.12688/f1000research.29032.2}
}

@inproceedings{babuji2019parsl,
  author    = {Babuji, Yadu and Woodard, Anna and Li, Zhuozhao and Katz,
               Daniel S. and Clifford, Ben and Kumar, Rohan and Lacinski,
               Lukasz and Chard, Ryan and Wozniak, Justin M. and Foster, Ian
               and Wilde, Michael and Chard, Kyle},
  title     = {Parsl: Pervasive parallel programming in {Python}},
  booktitle = {Proceedings of the 28th International Symposium on
               High-Performance Parallel and Distributed Computing (HPDC)},
  pages     = {25--36},
  year      = {2019},
  doi       = {10.1145/3307681.3325400}
}

@article{crusoe2022cwl,
  author  = {Crusoe, Michael R. and Abeln, Sanne and Iosup, Alexandru and
             Amstutz, Peter and Chilton, John and Tijani{\'c}, Neboj{\v s}a
             and M{\'e}nager, Herv{\'e} and Soiland-Reyes, Stian and
             Gavrilovi{\'c}, Bogdan and Goble, Carole and {The CWL
             Community}},
  title   = {Methods included: standardizing computational reuse and
             portability with the {Common Workflow Language}},
  journal = {Communications of the ACM},
  volume  = {65},
  number  = {6},
  pages   = {54--63},
  year    = {2022},
  doi     = {10.1145/3486897}
}

@article{basney2014cilogon,
  author  = {Basney, Jim and Fleury, Terry and Gaynor, Jeff},
  title   = {{CILogon}: A federated {X.509} certification authority for
             cyberinfrastructure logon},
  journal = {Concurrency and Computation: Practice and Experience},
  volume  = {26},
  number  = {13},
  pages   = {2225--2239},
  year    = {2014},
  doi     = {10.1002/cpe.3265}
}

@article{hudak2018openondemand,
  author  = {Hudak, David and Johnson, Doug and Chalker, Alan and Nicklas,
             Jeremy and Franz, Eric and Dockendorf, Trey and McMichael,
             Brian L.},
  title   = {{Open OnDemand}: A web-based client portal for {HPC} centers},
  journal = {Journal of Open Source Software},
  volume  = {3},
  number  = {25},
  pages   = {622},
  year    = {2018},
  doi     = {10.21105/joss.00622}
}

@article{mclennan2010hubzero,
  author  = {McLennan, Michael and Kennell, Rick},
  title   = {{HUBzero}: A platform for dissemination and collaboration in
             computational science and engineering},
  journal = {Computing in Science \& Engineering},
  volume  = {12},
  number  = {2},
  pages   = {48--53},
  year    = {2010},
  doi     = {10.1109/MCSE.2010.41}
}

@inproceedings{kluyver2016jupyter,
  author    = {Kluyver, Thomas and Ragan-Kelley, Benjamin and P{\'e}rez,
               Fernando and Granger, Brian and Bussonnier, Matthias and
               Frederic, Jonathan and Kelley, Kyle and Hamrick, Jessica and
               Grout, Jason and Corlay, Sylvain and Ivanov, Paul and Avila,
               Dami{\'a}n and Abdalla, Safia and Willing, Carol and
               {Jupyter Development Team}},
  title     = {Jupyter Notebooks: a publishing format for reproducible
               computational workflows},
  booktitle = {Positioning and Power in Academic Publishing: Players, Agents
               and Agendas (Proceedings of ELPUB 2016)},
  pages     = {87--90},
  year      = {2016},
  publisher = {IOS Press},
  doi       = {10.3233/978-1-61499-649-1-87}
}

@article{lawrence2015gateways,
  author  = {Lawrence, Katherine A. and Zentner, Michael and Wilkins-Diehr,
             Nancy and Wernert, Julie A. and Pierce, Marlon and Marru,
             Suresh and Michael, Scott},
  title   = {Science gateways today and tomorrow: positive perspectives of
             nearly 5000 members of the research community},
  journal = {Concurrency and Computation: Practice and Experience},
  volume  = {27},
  number  = {16},
  pages   = {4252--4268},
  year    = {2015},
  doi     = {10.1002/cpe.3526}
}

@article{baldin2019fabric,
  author  = {Baldin, Ilya and Nikolich, Anita and Griffioen, James and
             Monga, Indermohan Inder S. and Wang, Kuang-Ching and Lehman,
             Tom and Ruth, Paul},
  title   = {{FABRIC}: A national-scale programmable experimental network
             infrastructure},
  journal = {IEEE Internet Computing},
  volume  = {23},
  number  = {6},
  pages   = {38--47},
  year    = {2019},
  doi     = {10.1109/MIC.2019.2958545}
}

@inproceedings{gunter2011stampede,
  author    = {Gunter, Dan and Deelman, Ewa and Samak, Taghrid and Brooks,
               Christopher and Goode, Monte and Juve, Gideon and Mehta,
               Gaurang and Moraes, Priscilla and Silva, Fabio and Swany,
               Martin and Vahi, Karan},
  title     = {Online workflow management and performance analysis with
               {Stampede}},
  booktitle = {7th International Conference on Network and Service
               Management (CNSM 2011)},
  year      = {2011},
  publisher = {IEEE},
  doi       = {10.1109/CNSM.2011.6103947}
}

@inproceedings{vahi2012monitoring,
  author    = {Vahi, Karan and Harvey, Ian and Samak, Taghrid and Gunter, Dan
               and Evans, Kieran and Rogers, David and Taylor, Ian and Goode,
               Monte and Silva, Fabio and Al-Shakarchi, Eddie and Mehta,
               Gaurang and Jones, Andrew and Deelman, Ewa},
  title     = {A general approach to real-time workflow monitoring},
  booktitle = {7th Workshop on Workflows in Support of Large-Scale Science
               (WORKS 2012)},
  year      = {2012},
  publisher = {IEEE},
  doi       = {10.1109/SC.Companion.2012.21}
}

@misc{accesspegasus,
  author       = {{The Pegasus Team}},
  title        = {{ACCESS Pegasus}: a workflow science gateway for the
                  {NSF ACCESS} program},
  howpublished = {\url{https://pegasus.access-ci.org}},
  year         = {2026},
  note         = {Accessed 2026-10-15}
}

@misc{pegasusai-plugin,
  author       = {{The Pegasus Team}},
  title        = {{Pegasus-AI} plugin marketplace: skills, agents, and
                  knowledge base for {AI}-assisted {Pegasus} workflow
                  development},
  howpublished = {\url{https://github.com/pegasus-isi/claude-plugin-marketplace}},
  year         = {2026},
  note         = {Accessed 2026-10-15}
}

@misc{pegasusai-workflows,
  author       = {{The Pegasus Team}},
  title        = {{Pegasus} workflow repositories built with the
                  {Pegasus-AI} toolchain},
  howpublished = {\url{https://github.com/pegasus-isi}, repositories
                  \texttt{soilmoisture-workflow}, \texttt{airquality-workflow},
                  \texttt{earthquake-workflow}, \texttt{crophealth-workflow},
                  \texttt{mag-workflow}, \texttt{cper-soilmoisture-workflow},
                  \texttt{gw-workflow},
                  \texttt{eeg-workflow},
                  \texttt{casa-nowcast-workflow}, and others},
  year         = {2026},
  note         = {Accessed 2026-10-16}
}

@misc{eht-workflow,
  author       = {Thareja, Komal},
  title        = {Clean-room {Pegasus} reproduction of the {EHT M87}
                  imaging pipelines},
  howpublished = {\url{https://github.com/kthare10/eht-workflow}},
  year         = {2026},
  note         = {Accessed 2026-10-28}
}

@misc{pegasusai-ood,
  author       = {{The Pegasus Team}},
  title        = {{PegasusAI Studio} for {Open OnDemand}: a batch-connect
                  application},
  howpublished = {\url{https://github.com/pegasus-isi/bc_pegasus_ai_studio}},
  year         = {2026},
  note         = {Accessed 2026-10-16}
}

@misc{pegasusai-studio,
  author       = {{The Pegasus Team}},
  title        = {{PegasusAI Studio}: an {AI}-assisted science gateway for
                  {Pegasus} workflows},
  howpublished = {Source code at \url{https://github.com/pegasus-isi/studio-webapp};
                  public gateway at \url{https://studio.pegasus-ai.org};
                  project site at \url{https://pegasus-ai.org}},
  year         = {2026},
  note         = {Accessed 2026-10-15}
}

@article{galaxy2022,
  author  = {{The Galaxy Community}},
  title   = {The {Galaxy} platform for accessible, reproducible and
             collaborative biomedical analyses: 2022 update},
  journal = {Nucleic Acids Research},
  volume  = {50},
  number  = {W1},
  pages   = {W345--W351},
  year    = {2022},
  doi     = {10.1093/nar/gkac247}
}

@article{boiko2023coscientist,
  author  = {Boiko, Daniil A. and MacKnight, Robert and Kline, Ben and
             Gomes, Gabe},
  title   = {Autonomous chemical research with large language models},
  journal = {Nature},
  volume  = {624},
  number  = {7992},
  pages   = {570--578},
  year    = {2023},
  doi     = {10.1038/s41586-023-06792-0}
}

@article{bran2024chemcrow,
  author  = {M. Bran, Andres and Cox, Sam and Schilter, Oliver and Baldassari,
             Carlo and White, Andrew D. and Schwaller, Philippe},
  title   = {Augmenting large language models with chemistry tools},
  journal = {Nature Machine Intelligence},
  volume  = {6},
  number  = {5},
  pages   = {525--535},
  year    = {2024},
  doi     = {10.1038/s42256-024-00832-8}
}

@misc{mcp2024,
  author       = {{Anthropic}},
  title        = {Model Context Protocol},
  year         = {2024},
  howpublished = {\url{https://modelcontextprotocol.io}},
  note         = {Accessed 2026-08-11}
}

@inproceedings{hancock2021jetstream2,
  author    = {Hancock, David Y. and Fischer, Jeremy and Lowe, John Michael
               and Snapp-Childs, Winona and Pierce, Marlon and Marru, Suresh
               and Coulter, J. Eric and Vaughn, Matthew and Beck, Brian and
               Merchant, Nirav and Skidmore, Edwin and Jacobs, Gwen},
  title     = {Jetstream2: Accelerating cloud computing via {Jetstream}},
  booktitle = {Practice and Experience in Advanced Research Computing
               (PEARC '21)},
  pages     = {1--8},
  year      = {2021},
  doi       = {10.1145/3437359.3465565}
}

@article{patel2023eht,
  author  = {Patel, Ria and Roachell, Brandan and Ca{\'i}{\~n}o-Lores, Silvina and Ketron, Ross and Leonard, Jacob and Tan, Nigel and Vahi, Karan and Brown, Duncan A. and Deelman, Ewa and Taufer, Michela},
  title   = {Reproducibility of the First Image of a Black Hole in the Galaxy {M87} From the Event Horizon Telescope Collaboration},
  journal = {Computing in Science \& Engineering},
  volume  = {24},
  number  = {5},
  pages   = {42--52},
  year    = {2022},
  doi     = {10.1109/MCSE.2023.3241105}
}

@article{ehtIV2019,
  author  = {{Event Horizon Telescope Collaboration} and others},
  title   = {First {M87} {Event Horizon Telescope} Results. {IV}. Imaging
             the Central Supermassive Black Hole},
  journal = {The Astrophysical Journal Letters},
  volume  = {875},
  number  = {1},
  pages   = {L4},
  year    = {2019},
  doi     = {10.3847/2041-8213/ab0e85}
}

@article{abbott2019gwtc1,
  author  = {Abbott, B.~P. and others},
  title   = {{GWTC-1}: A Gravitational-Wave Transient Catalog of Compact
             Binary Mergers Observed by {LIGO} and {Virgo} during the
             First and Second Observing Runs},
  journal = {Physical Review X},
  volume  = {9},
  number  = {3},
  pages   = {031040},
  year    = {2019},
  doi     = {10.1103/PhysRevX.9.031040}
}

@misc{pegasus_method_companion,
  author = {{PegasusAI Team}},
  title = {Napkin to Science: From Scientific Intent to Executable Workflows on Production Cyberinfrastructure},
  year = {2026},
  note = {Under submission to Future Generation Computer Systems}
}

@techreport{biggs2023aigateways,
  author      = {Biggs, Jr., Brandon Samuel},
  title       = {Integrating Artificial Intelligence into Science Gateways},
  institution = {Idaho National Laboratory},
  number      = {INL/CON-23-72098},
  year        = {2023},
  url         = {https://inldigitallibrary.inl.gov/sites/sti/sti/Sort_65663.pdf}
}

@inproceedings{sperhac2024aigateways,
  author    = {Sperhac, Jeanette and Gesing, Sandra and Zentner, Michael and
               Stirm, Claire and Quick, Robert and Stubbs, Joe},
  title     = {{AI} and Science Gateways: A Promising Combination for
               Accelerating Science and Research Computing},
  booktitle = {Practice and Experience in Advanced Research Computing
               (PEARC '24)},
  publisher = {ACM},
  year      = {2024},
  doi       = {10.1145/3626203.3670562}
}

@misc{ma2025llmhpc,
  author        = {Ma, Heng and Brace, Alexander and Siebenschuh, Carlo and
                   Pauloski, Greg and Foster, Ian and Ramanathan, Arvind},
  title         = {Connecting Large Language Model Agent to High Performance
                   Computing Resource},
  year          = {2025},
  eprint        = {2502.12280},
  archivePrefix = {arXiv},
  primaryClass  = {cs.DC},
  doi           = {10.48550/arXiv.2502.12280}
}

@article{mclaughlin2009casa,
  author  = {McLaughlin, David and Pepyne, David and Chandrasekar, V. and Philips, Brenda and Kurose, James and Zink, Michael and Droegemeier, Kelvin and Cruz-Pol, Sandra and Junyent, Francesc and Brotzge, Jerald and Westbrook, David and Bharadwaj, Nitin and Wang, Yanting and Lyons, Eric and Hondl, Kurt and Liu, Yuxiang and Knapp, Eric and Xue, Ming and Hopf, Anthony and Kloesel, Kevin and DeFonzo, Alfred and Kollias, Pavlos and Brewster, Keith and Contreras, Robert and Djaferis, Theodore and Insanic, Edin and Frasier, Stephen and Carr, Frederick},
  title   = {Short-Wavelength Technology and the Potential for Distributed Networks of Small Radar Systems},
  journal = {Bulletin of the American Meteorological Society},
  volume  = {90},
  number  = {12},
  pages   = {1797--1818},
  year    = {2009},
  doi     = {10.1175/2009BAMS2507.1}
}

@article{abbott2021gwosc,
  author  = {Abbott, R. and others},
  title   = {Open Data from the First and Second Observing Runs of
             {Advanced LIGO} and {Advanced Virgo}},
  journal = {SoftwareX},
  volume  = {13},
  pages   = {100658},
  year    = {2021},
  doi     = {10.1016/j.softx.2021.100658}
}

@article{markiewicz2021openneuro,
  author  = {Markiewicz, Christopher J. and Gorgolewski, Krzysztof J. and Feingold, Franklin and Blair, Ross and Halchenko, Yaroslav O. and Miller, Eric and Hardcastle, Nell and Wexler, Joe and Esteban, Oscar and Goncalves, Mathias and Jwa, Anita and Poldrack, Russell},
  title   = {The {OpenNeuro} Resource for Sharing of Neuroscience Data},
  journal = {eLife},
  volume  = {10},
  pages   = {e71774},
  year    = {2021},
  doi     = {10.7554/eLife.71774}
}

@article{kurtzer2017singularity,
  author  = {Kurtzer, Gregory M. and Sochat, Vanessa and Bauer, Michael W.},
  title   = {Singularity: Scientific containers for mobility of compute},
  journal = {PLOS ONE},
  volume  = {12},
  number  = {5},
  pages   = {e0177459},
  year    = {2017},
  doi     = {10.1371/journal.pone.0177459}
}

@inproceedings{yoo2003slurm,
  author    = {Yoo, Andy B. and Jette, Morris A. and Grondona, Mark},
  title     = {{SLURM}: Simple {Linux} Utility for Resource Management},
  booktitle = {Job Scheduling Strategies for Parallel Processing (JSSPP 2003)},
  series    = {Lecture Notes in Computer Science},
  volume    = {2862},
  pages     = {44--60},
  publisher = {Springer},
  year      = {2003},
  doi       = {10.1007/10968987_3}
}

@article{pordes2007osg,
  author  = {Pordes, Ruth and Petravick, Don and Kramer, Bill and Olson, Doug
             and Livny, Miron and Roy, Alain and Avery, Paul and Blackburn,
             Kent and Wenaus, Torre and W{\"u}rthwein, Frank and others},
  title   = {The {Open Science Grid}},
  journal = {Journal of Physics: Conference Series},
  volume  = {78},
  pages   = {012057},
  year    = {2007},
  doi     = {10.1088/1742-6596/78/1/012057}
}

@inproceedings{strande2021expanse,
  author    = {Strande, Shawn and Cai, Haisong and Tatineni, Mahidhar and
               Pfeiffer, Wayne and Irving, Christopher and Majumdar, Amit and
               Mishin, Dmitry and Sinkovits, Robert and Norman, Mike and
               Wolter, Nicole and others},
  title     = {Expanse: Computing without Boundaries},
  booktitle = {Practice and Experience in Advanced Research Computing
               (PEARC '21)},
  publisher = {ACM},
  pages     = {1--4},
  year      = {2021},
  doi       = {10.1145/3437359.3465588}
}

@article{burns2016kubernetes,
  author  = {Burns, Brendan and Grant, Brian and Oppenheimer, David and
             Brewer, Eric and Wilkes, John},
  title   = {Borg, {Omega}, and {Kubernetes}},
  journal = {Queue},
  volume  = {14},
  number  = {1},
  pages   = {70--93},
  year    = {2016},
  doi     = {10.1145/2898442.2898444}
}

@misc{kimi2025k2,
  author        = {{Kimi Team}},
  title         = {{Kimi K2}: Open Agentic Intelligence},
  year          = {2025},
  eprint        = {2507.20534},
  archivePrefix = {arXiv},
  primaryClass  = {cs.LG}
}

@misc{qwen2025qwen3,
  author        = {Yang, An and others},
  title         = {{Qwen3} Technical Report},
  year          = {2025},
  eprint        = {2505.09388},
  archivePrefix = {arXiv},
  primaryClass  = {cs.CL}
}

@misc{tool-claudecode,
  author = {{Anthropic}}, title = {{Claude Code}},
  howpublished = {\url{https://github.com/anthropics/claude-code}},
  year = {2026}, note = {Accessed 2026-10-29}
}

@misc{tool-opencode,
  author = {{OpenCode contributors}}, title = {{OpenCode}},
  howpublished = {\url{https://github.com/anomalyco/opencode}},
  year = {2026}, note = {Accessed 2026-10-29}
}

@misc{tool-crush,
  author = {{Charm}}, title = {{Crush}},
  howpublished = {\url{https://github.com/charmbracelet/crush}},
  year = {2026}, note = {Accessed 2026-10-29}
}

@misc{tool-codex,
  author = {{OpenAI}}, title = {{Codex CLI}},
  howpublished = {\url{https://github.com/openai/codex}},
  year = {2026}, note = {Accessed 2026-10-29}
}

@misc{tool-antigravity,
  author = {{Google}}, title = {{Antigravity}},
  howpublished = {\url{https://antigravity.google}},
  year = {2026}, note = {Accessed 2026-10-29}
}

@misc{opentelemetry,
  author = {{OpenTelemetry Authors}}, title = {{OpenTelemetry}},
  howpublished = {\url{https://opentelemetry.io}},
  year = {2026}, note = {Accessed 2026-10-29}
}

@misc{kiso,
  author = {{The Pegasus Team}},
  title  = {{Kiso}: experiment management for {Pegasus} workflows on cloud and edge testbeds},
  howpublished = {\url{https://github.com/pegasus-isi/kiso}},
  year = {2026}, note = {Accessed 2026-10-29}
}

@misc{nrp,
  author       = {{National Research Platform}},
  title        = {{National Research Platform (NRP)}},
  howpublished = {\url{https://nrp.ai}},
  year         = {2026},
  note         = {Accessed 2026-10-29}
}

@techreport{biggs2024hpcassistant,
  author      = {Biggs, Brandon S. and Dalton, Kaylee},
  title       = {Supplementing {HPC} Support with a Science Gateway {AI} Assistant},
  institution = {Idaho National Laboratory},
  number      = {INL/CON-24-76939},
  year        = {2024},
  month       = may,
  url         = {https://inldigitallibrary.inl.gov/content/uploads/50/2026/04/Sort_93465.pdf}
}

@article{merkel2014docker,
  author  = {Merkel, Dirk},
  title   = {Docker: Lightweight {Linux} Containers for Consistent Development
             and Deployment},
  journal = {Linux Journal},
  volume  = {2014},
  number  = {239},
  pages   = {2},
  year    = {2014}
}
